\documentclass[aps,prd,reprint,amsmath,amssymb,superscriptaddress,floatfix]{revtex4}
\usepackage{graphicx}
\usepackage{amsfonts}
\usepackage{amssymb}
\usepackage{rotating}
\usepackage{booktabs}
\usepackage{xcolor}
\usepackage{soul}
\usepackage{color}
\usepackage{slashed}
\usepackage{multirow}
\usepackage{makecell}
\usepackage{epsf}
\usepackage{ulem}
\usepackage{cancel}
\usepackage{color,bm}
\usepackage{bm}
\usepackage{mathtools}

\usepackage[colorinlistoftodos]{todonotes}

\usepackage[colorlinks=true,citecolor=cyan,urlcolor=blue,bookmarks=true,bookmarks=true,bookmarksopen=true,bookmarksnumbered=true,bookmarksopenlevel=3]{hyperref}

\definecolor{airforceblue}{rgb}{0.36, 0.54, 0.66}
\definecolor{steelblue}{rgb}{0.27, 0.51, 0.71}
\definecolor{amber}{rgb}{1.0, 0.49, 0.0}

\newcommand{\s}{\slashed}

\begin{document}

\title{ T-odd generalized and quasi transverse momentum dependent parton distribution in a scalar spectator model }
\author{Xuan Luo}
\author{\textsc{Hao Sun}\footnote{Corresponding author: haosun@mail.ustc.edu.cn \hspace{0.2cm} haosun@dlut.edu.cn}}
\affiliation{ Institute of Theoretical Physics, School of Physics, Dalian University of Technology, \\ No.2 Linggong Road, Dalian, Liaoning, 116024, P.R.China }
\date{\today}

\begin{abstract}
Generalized transverse momentum dependent parton distributions (GTMDs), 
as mother funtions of transverse momentum dependent parton distributions (TMDs) 
and generalized parton distributions (GPDs), encode the most general parton structure of hadrons.
We calculate four twist-two time reversal odd GTMDs of pion in a scalar spectator model.
We study the dependence of GTMDs on the longitudinal momentum fraction $x$ carried by the active quark 
and the transverse momentum $|\vec k_T|$ for different values of skewness $\xi$ defined as the longitudinal momentum transferred to the proton
as well as the total momentum $|\vec\Delta_T|$ transferred to the proton.
In addition, the quasi-TMDs and quasi-GPDs of pion have also been investigated in this paper.
\vspace{0.5cm}
\end{abstract}
\maketitle
\setcounter{footnote}{0}

\section{INTRODUCTION}
\label{I}

Exploring the partonic substructure of hadrons is still at the frontier of hadronic high energy physics research. 
The parton distribution functions (PDFs) make it clear how the longitudinal parton momentums in hadron are distributed. 
However, they only include one dimension information. Therefore probing how partons distribute 
in the transverse plane in both momentum and coordinate space becomes a vital topic. 
Typically, the transverse spatial distribution of partons inside a hadron can be quantified 
by Generalised parton distributions (GPDs) \cite{Diehl:2003ny,Belitsky:2005qn,Garcon:2002jb}, 
which can be denoted as a function of longitudinal momentum fraction $x$ carried by the parton, 
the longitudinal momentum $\xi$ transferred to the hadron and the total momentum transferred $t$.
They can be accessed through measurements in hard exclusive reactions like deep virtual Compton scattering 
and hard exclusive meson production \cite{Goeke:2001tz,Belitsky:2005qn}.
While the transverse momentum dependent parton distributions (TMDs) \cite{Collins:1981uw,Bacchetta:2006tn,Meissner:2007rx} 
depending on both the longitudinal and transverse motion of partons inside a hadron 
can be studied by the description of various hard semi-inclusive reactions.
More general distibutions than the GPDs and the TMDs, the generalized transverse momentum dependent 
parton distributions (GTMDs) \cite{Meissner:2008ay,Meissner:2009ww,Lorce:2013pza} 
could reduce to them in specific kinematical limits, therefore serve as mother distributions. 
The GTMDs can directly enter the description of hard exclusive reactions.
The parametrization of the generalized quark-quark correlation functions for a spin-$0$ and spin-$1/2$ hadron 
in terms of GTMDs are given in Refs.\cite{Meissner:2008ay,Meissner:2009ww}. 
Then the authors in Ref.\cite{Lorce:2013pza} add a complete classification of gluon GTMDs.
Particularly, the correlator related to the time-reversal odd (T-odd) GTMDs 
is contributed by the final state interactions from gauge link or Wilson line. 
These interactions are necessary to generate the single spin asymmetries \cite{Brodsky:2002cx}.

Although PDFs are related to parton fields in QCD, it is difficult to calculate them directly in QCD since they are nonperturbative quantities. 
This difficulties may be overcomed by Lattice QCD method studying the PDFs from first principles. 
However, PDFs are usually defined on the light cone, which poses a problem for the standard Euclidean formulation, 
and in lattice QCD calculation only moments of distributions in $x$ can be accessed as matrix elements of local operators \cite{Dolgov:2002zm,Gockeler:2004wp}. 
To overcome these issues, the proposed large-momentum effective theory (LaMET) of Ji has been presented \cite{Ji:2013dva}. 
This method evaluates PDFs on the lattice through quasi-PDFs \cite{Ji:2013dva,Ma:2014jla,Ji:2014gla}, 
whose mother correlation functions includes a spacelike operator $\gamma^z$ 
instead of the usual lightlike $\gamma^+$ entering the definition of the standard PDFs. 
These quasi-PDFs can be reached directly from the lattice QCD calculation \cite{Lin:2014zya} 
and as the quasi-PDFs depend logarithmically on $P_z$ when $P_z$ becomes large, 
and they need a perturbative matching in LaMET to reduce to the standard PDFs. 
A very recent review of LaMET is given in \cite{Ji:2020ect}.
Many theoretical discussions and lattice simulations for quasi-PDFs and similar quantities 
has been performed \cite{Orginos:2017kos,Green:2017xeu,Alexandrou:2015rja,Chen:2016utp,Alexandrou:2016jqi,Zhang:2017bzy,Lin:2017ani,Bali:2017gfr,Alexandrou:2017dzj,Chen:2017gck,Alexandrou:2018pbm,Chen:2018xof,Chen:2018fwa,Alexandrou:2018eet,Liu:2018uuj,Lin:2018pvv}. 
Moreover, several model calculations of quasi-PDFs have been carried out 
\cite{Gamberg:2014zwa,Bacchetta:2016zjm,Nam:2017gzm,Broniowski:2017wbr,Hobbs:2017xtq,Broniowski:2017gfp,Xu:2018eii}.
There have also been a number of works on quasi-PDFs renormalization 
\cite{Chen:2017mzz,Green:2017xeu,Ji:2017oey,Ishikawa:2017faj,Alexandrou:2017huk,Constantinou:2017sej,Xiong:2017jtn,Chen:2016fxx,Ishikawa:2016znu}.
The approach \cite{Ji:2013dva} can be generalized to any light-cone correlations in hadron physics, e.g. the correlators related to GPDs and TMDs.
There has been a lot of efforts on the quasi-TMDs, 
including their renormalization and matching to the physical TMDs 
\cite{Ji:2014hxa,Ji:2018hvs,Ebert:2018gzl,Ebert:2019okf,Ebert:2019tvc,Ji:2019sxk,Ji:2019ewn,Vladimirov:2020ofp,Ji:2020ect,Ebert:2020gxr}. 
These quasi-TMD works have made important breakthroughs on their pinched-pole singularity issue, 
renormalization, evolution, soft factor subtraction, and factorization into physical TMDs.
Moreover, the nonperturbative Collins-Soper evolution kernel of TMDs has been calculated in 
\cite{Shanahan:2019zcq,Shanahan:2020zxr} , 
which is an important step in TMDs full extractions from lattice QCD.
In summary, it may be useful to study the quasi-distributions such as quasi-PDFs, quasi-GPDs and quasi-TMDs.

Among the hadrons, pions are very fascinating particles and they hold a lot of information on the structure of hadrons. 
There has been a tremendous effort to deduce the parton distribution functions of the pion.
Pions provide the force that binds the protons and neutrons inside the nuclei and they also influence the properties of the isolated nucleons. 
Thus understanding of matter is not complete without getting a detailed information on the role of pions.
In this paper, being inspired by the previous works for quasi-distribution model calculations \cite{Bhattacharya:2018zxi,Bhattacharya:2019cme,Ma:2019agv}, 
we will probe the T-odd GTMDs, quasi-TMDs and the quasi-GPDs of the pion applying a scalar spectator model.
In particular, GPDs of the pion have been obtained in various models like chiral quark model \cite{Broniowski:2003rp,Dorokhov:2011ew}, 
NJL model \cite{Davidson:2001cc,Theussl:2002xp}, light-front constituent quark model \cite{Frederico:2009fk} 
and lattice QCD \cite{Brommel:2007xd,Sufian:2020vzb,Izubuchi:2019lyk,Chen:2019lcm}.
We will give out the analytical results of all four twist-two T-odd GTMDs, quasi-TMDs and quasi-GPDs in the present paper, 
and conduct a qualitative analysis of all these distributions.

The remainder of this paper is as follows: 
Sec.\ref{II} below describes in detail the theoretical definition of various pion parton distributions. 
In Sec.\ref{III} we give out the analytical results of four T-odd GTMDs, quasi-TMDs and quasi-GPDs in a scalar spectator. 
In Sec.\ref{IV}, we present our numerical studies using a group of fitted model parameters of a scalar spectator model. 
A brief conclusion is presented in Sect.\ref{V}.

\section{Definition of pion parton distributions}
\label{II}

\subsection{Transverse momentum dependent parton distribution $h_{1\pi}^{\perp}$}

For a spinless particle, such as the pion, only two leading twist TMDs arise, in contrast to the eight found for spin-$\frac{1}{2}$ particles \cite{Barone:2010zz}.
The TMD $f_{1\pi}$ is simply the unpolarized quark distribution, whereas the Boer-Mulders (BM) function \cite{Boer:1997nt}, $h_{1\pi}^{\perp}$, 
describes the distribution of transversely polarized quarks in the pion.
The BM function $h_{1\pi}^{\perp}$ is defined from the quark-quark distribution correlation function 
\begin{eqnarray} \label{eq1}
\begin{aligned}
\Phi^{[\Gamma]}(x,\vec{k}_T) = \int \frac{1}{2}\frac{d\xi^- d^2 \vec{\xi}_T}{(2\pi)^3}e^{ik \cdot \xi} 
\langle P|\bar{\psi}(-\frac{1}{2}\xi)\Gamma\mathcal{W}_{+\infty}(-\frac{1}{2}\xi;\frac{1}{2}\xi)\psi(\frac{1}{2}\xi)|P \rangle \bigg|_{\xi^+=0},
\end{aligned}
\end{eqnarray}
where $P$ is the four-momentum of the pion moving along the $z$-axis with the components ($P^+,P^-,\vec{0}_\perp$) in the light-cone coordinates, 
in which the plus and minus components of any four-vector $a^\mu$ have the form $a^\pm=(a^0\pm a^3)/\sqrt{2}$, and transverse part $\vec{a}_\perp=(a^1,a^2)$.
The quark field and momentum are denoted by $k$ and $\psi$.
In the correlator Eq.(\ref{eq1}) we have Wilson lines
\begin{eqnarray} \label{eq2}
\begin{aligned}
\mathcal{W}_{+\infty}(-\frac{1}{2}\xi;\frac{1}{2}\xi)|_{\xi^+=0}&=[0^+,-\frac{1}{2}\xi^-,-\frac{1}{2}\vec{\xi}_T;0^+,+\infty^-,-\frac{1}{2}\vec{\xi}_T]
\cdot [0^+,+\infty^-,-\frac{1}{2}\vec{\xi}_T;0^+,+\infty^-,\frac{1}{2}\vec{\xi}_T] 
\\
&\cdot [0^+,+\infty^-,\frac{1}{2}\vec{\xi}_T;0^+,\frac{1}{2}\xi^-,\frac{1}{2}\vec{\xi}_T],
\end{aligned}
\end{eqnarray}
where
\begin{eqnarray} \label{eq3}
\begin{aligned}
\big[0^+,-\frac{1}{2}\xi^-,-\frac{1}{2}\vec{\xi}_T;0^+,+\infty^-,-\frac{1}{2}\vec{\xi}_T\big]&=\mathcal{P}\exp\bigg[ -ig_s\int_{-\frac{1}{2}\xi^-}^{+\infty^-} d\zeta A^+(\zeta^-,0^+,-\frac{1}{2}\vec{\xi}_T) \bigg],
\\
[0^+,+\infty^-,-\frac{1}{2}\vec{\xi}_T;0^+,+\infty^-,\frac{1}{2}\vec{\xi}_T]&=\mathcal{P}\exp\bigg[ -ig_s\int_{-\frac{1}{2}\vec{\xi}_T}^{\frac{1}{2}\vec{\xi}_T} d\zeta _T A_T(+\infty^-,0^+,\vec{\zeta}_T) \bigg],
\\
[0^+,+\infty^-,\frac{1}{2}\vec{\xi}_T;0^+,\frac{1}{2}\xi^-,\frac{1}{2}\vec{\xi}_T]&=\mathcal{P}\exp\bigg[ -ig_s\int_{+\infty^-}^{\frac{1}{2}\xi^-} d\zeta A^+(\zeta^-,0^+,\frac{1}{2}\vec{\xi}_T) \bigg].
\end{aligned}
\end{eqnarray}
Here $\mathcal{P}$ is the path-ordering operator and $A$ is the gluon field.
The strong couping constant is denoted by $g_s$.
The BM function $h_{1\pi}^{\perp}$ can be obtained by
\begin{eqnarray} \label{eq4}
\begin{aligned}
\Phi^{[i\sigma^{\alpha+}\gamma_5]}=-\frac{\varepsilon_T^{\alpha\beta}k_T^\beta}{M}h_{1\pi}^\perp ,
\end{aligned}
\end{eqnarray}
with the pion mass denoted by $M$. 

\subsection{T-odd GTMDs and GPDs}

To obtain GTMDs, we start from the generalized $k_T$-dependent correlator denoted by
\begin{eqnarray} \label{eq5}
\begin{aligned}
W^{[\Gamma]}(x,\vec{k}_T,\Delta)=\int \frac{1}{2}\frac{d\xi^- d^2 \vec{\xi}_T}{(2\pi)^3}e^{ik \cdot \xi} \langle p'|\bar{\psi}(-\frac{1}{2}\xi)\Gamma\mathcal{W}_{+\infty}(-\frac{1}{2}\xi;\frac{1}{2}\xi)\psi(\frac{1}{2}\xi)|p \rangle \bigg|_{\xi^+=0},
\end{aligned}
\end{eqnarray}
where the initial and final state four-momentum are characterized by $p$ and $p'$.
We use the common kinematical variables
\begin{eqnarray} \label{eq6}
\begin{aligned}
P&=\frac{1}{2}(p+p'), \qquad \Delta=p'-p, \qquad t=\Delta^2=-\frac{1}{1-\xi^2}(4\xi^2 M^2+\vec \Delta_T^2),
\\
P^\mu&=\left[ P^+,\frac{4M^2+\vec \Delta_T^2}{8(1-\xi^2)P^+},\vec{0}_T \right], \qquad k^\mu=\left[ xP^+,\frac{k^2+\vec k_T^2}{2xP^+},\vec{k}_T \right],
\\
\Delta^\mu&=\left[ -2\xi P^+,\frac{\xi(4 M^2+\vec \Delta_T^2)}{4(1-\xi^2)P^+},\vec{\Delta}_T \right],
\end{aligned}
\end{eqnarray}
where we consider the range $0 \le \xi \le 1$ of the skewness variable $\xi$. 
In general, the generalized $k_T$-dependent correlator in Eq.(\ref{eq5}), unlike GPDs or TMDs, are complex-valued functions.
We can reach four complex-valued twist-two GTMDs $F_1,\tilde{G}_1,H_1^k,H_1^\Delta$ through
\begin{eqnarray} \label{eq7}
\begin{aligned}
W^{[\gamma^+]}&=F_1^e+iF_1^o,
\\
W^{[\gamma^+\gamma_5]}&=\frac{i\varepsilon_T^{ij} k_T^i \Delta_T^j}{M^2}(\tilde{G}_1^e+i\tilde{G}_1^o),
\\
W^{[i\sigma^{j+}\gamma_5]}&=\frac{i\varepsilon_T^{ij} k_T^i}{M}(H_1^{k,e}+iH_1^{k,o})+\frac{i\varepsilon_T^{ij} \Delta_T^i}{M}(H_1^{\Delta,o}+iH_1^{\Delta,o}),
\end{aligned}
\end{eqnarray}
where the superscripts $e$, $o$ stand for T-even and T-odd part respectively.
We have adopted the general definitions $\sigma^{\mu\nu}=i[\gamma^\mu,\gamma^\nu]/2$, $\varepsilon^{0123}=1$ and $\varepsilon_T^{ij}=\varepsilon^{-+ij}$.
The T-odd twist-two GTMDs correspond to the imaginary part of complex-valued twist-two GTMDs.

The twist-two standard GPDs of quarks for a spin-0 hadron come from the integrated quark-quark correlator obtained from the correlator $W$ in Eq.(\ref{eq5}) by means of the projection
\begin{eqnarray} \label{eq8}
\begin{aligned}
F^{[\Gamma]}(x,\Delta)=\int dk^- d^2 \vec{k}_T W^{[\Gamma]}(x,\vec{k}_T,\Delta)=\int \frac{1}{2}\frac{d\xi^-}{2\pi}e^{ik \cdot \xi} \langle p'|\bar{\psi}(-\frac{1}{2}\xi)\Gamma\mathcal{W}(-\frac{1}{2}\xi;\frac{1}{2}\xi)\psi(\frac{1}{2}\xi)|p \rangle \bigg|_{\xi^+=0} .
\end{aligned}
\end{eqnarray}
The GPDs parameterize the Dirac traces $F^{[\Gamma]}$ of the GPD-correlator in Eq.(\ref{eq8}) and there are only two GPDs in twist-two
\begin{eqnarray} \label{eq9}
\begin{aligned}
F^{[\gamma^+]}&=F_1(x,\xi,t),
\\
F^{[i\sigma^{j+}\gamma_5]}&=\frac{i\varepsilon_T^{ij}\Delta_T^i}{M}H_1(x,\xi,t) .
\end{aligned}
\end{eqnarray}

\subsection{Quasi-TMD and quasi-GPD}

In the following we turn to the definitions of the quasi-TMD and quasi-GPD of the pion meson.
Quasi-TMD $h_{1\pi}^{\perp}(x,\vec k_T^2;P_z)$ is defined through an equal-time spatial correlation function
\begin{eqnarray} \label{eq10}
\begin{aligned}
\Phi^{[\Gamma]}(x,\vec{k}_T;P_z)=\int \frac{1}{2}\frac{d\xi^- d^2 \vec{\xi}_T}{(2\pi)^3}e^{ik \cdot \xi} \langle P|\bar{\psi}(-\frac{1}{2}\xi)\Gamma\mathcal{W}_{Q,+\infty}(-\frac{1}{2}\xi;\frac{1}{2}\xi)\psi(\frac{1}{2}\xi)|P \rangle \bigg|_{\xi^+=0}.
\end{aligned}
\end{eqnarray}
The Wilson lines read
\begin{eqnarray} \label{eq11}
\begin{aligned}
&\mathcal{W}_{Q,+\infty}(-\frac{1}{2}\xi;\frac{1}{2}\xi)|_{\xi^+=0}=[0,-\frac{1}{2}\xi^3,-\frac{1}{2}\vec{\xi}_T;0,+\infty^3,-\frac{1}{2}\vec{\xi}_T]
\\
&\cdot [0,+\infty^3,-\frac{1}{2}\vec{\xi}_T;0,+\infty^3,\frac{1}{2}\vec{\xi}_T] \cdot [0,+\infty^3,\frac{1}{2}\vec{\xi}_T;0,\frac{1}{2}\xi^3,\frac{1}{2}\vec{\xi}_T],
\end{aligned}
\end{eqnarray}
where
\begin{eqnarray} \label{eq12}
\begin{aligned}
\big[0,-\frac{1}{2}\xi^3,-\frac{1}{2}\vec{\xi}_T;0,+\infty^3,-\frac{1}{2}\vec{\xi}_T\big]&=\mathcal{P}\exp\bigg[ -ig_s\int_{-\frac{1}{2}\xi^3}^{+\infty^3} d\zeta A^3(\zeta^3,0,-\frac{1}{2}\vec{\xi}_T) \bigg],
\\
[0,+\infty^3,-\frac{1}{2}\vec{\xi}_T;0,+\infty^3,\frac{1}{2}\vec{\xi}_T]&=\mathcal{P}\exp\bigg[ -ig_s\int_{-\frac{1}{2}\vec{\xi}_T}^{\frac{1}{2}\vec{\xi}_T} d\zeta_T A_T(+\infty^3,0,\vec{\zeta}_T) \bigg],
\\
[0,+\infty^3,\frac{1}{2}\vec{\xi}_T;0,\frac{1}{2}\xi^3,\frac{1}{2}\vec{\xi}_T]&=\mathcal{P}\exp\bigg[ -ig_s\int_{+\infty^3}^{\frac{1}{2}\xi^3} d\zeta A^+(\zeta^3,0,\frac{1}{2}\vec{\xi}_T) \bigg].
\end{aligned}
\end{eqnarray}
Quasi-TMD $h_{1\pi,Q}^{\perp}(x,\vec k_T^2;P_z)$ can be obtained from two definitions
\begin{eqnarray} \label{eq13}
\begin{aligned}
&\Phi^{[i\sigma^{\alpha3}\gamma_5]}=-\frac{\varepsilon_T^{\alpha\beta}k_T^\beta}{M}h_{1\pi,Q(3)}^\perp(x,\vec k_T^2;P_z),
\\
&\Phi^{[i\sigma^{\alpha0}\gamma_5]}=-\frac{\varepsilon_T^{\alpha\beta}k_T^\beta}{M}h_{1\pi,Q(0)}^\perp(x,\vec k_T^2;P_z).
\end{aligned}
\end{eqnarray}
The original paper on quasi-PDFs suggested to use the matrix $\gamma^3$ \cite{Ji:2013dva} for the unpolarized quasi-PDF $f_{1,Q}(x;P^3)$. It was later argued that the matrix $\gamma^0$ would lead to a better suppression of higher-twist contributions \cite{Radyushkin:2016hsy}.
Similarly, quasi-GPDs are also defined through an equal-time spatial correlation function
\begin{eqnarray} \label{eq14}
\begin{aligned}
F_Q^{[\Gamma]}(x,\Delta;P_z)=\int \frac{1}{2}\frac{d\xi^-}{2\pi}e^{ik \cdot \xi} \langle p'|\bar{\psi}(-\frac{1}{2}\xi)\Gamma\mathcal{W}_{Q}(-\frac{1}{2}\xi;\frac{1}{2}\xi)\psi(\frac{1}{2}\xi)|p \rangle \bigg|_{\xi^0=0,\vec{z}_\perp=\vec{0}_\perp},
\end{aligned}
\end{eqnarray}
where the Wilson line is given by
\begin{eqnarray} \label{eq15}
\begin{aligned}
\mathcal{W}_{Q}(-\frac{1}{2}\xi;\frac{1}{2}\xi) \bigg|_{\xi^0=0,\vec{z}_\perp=\vec{0}_\perp}=\mathcal{P}\exp\bigg( -ig_s\int_{-\frac{\xi^3}{2}}^{\frac{\xi^3}{2}} dy^3 A^3(0,\vec{0}_\perp;y^3) \bigg).
\end{aligned}
\end{eqnarray}
Then the twist-two quasi-GPDs of the pion can be obtained through two definitions
\begin{eqnarray} \label{eq16}
\begin{aligned}
F^{[\gamma^3]}&=F_{1,Q(3)}(x,\xi,t;P_z), \qquad \qquad \qquad \ \ F^{[\gamma^0]}=F_{1,Q(0)}(x,\xi,t;P_z),
\\
F^{[i\sigma^{j3}\gamma_5]}&=\frac{i\varepsilon_T^{ij}\Delta_T^i}{M}H_{1,Q(3)}(x,\xi,t;P_z), \qquad F^{[i\sigma^{j0}\gamma_5]}=\frac{i\varepsilon_T^{ij}\Delta_T^i}{M}H_{1,Q(0)}(x,\xi,t;P_z).
\end{aligned}
\end{eqnarray}

\section{Analytical results in a scalar spectator model}
\label{III}

In this section, being inspired by the previous works for quasi-distribution model calculations \cite{Bhattacharya:2018zxi,Bhattacharya:2019cme,Ma:2019agv}, 
we apply a scalar spectator model to reach the analytic results of the quasi-TMD $h_{1\pi}^{\perp}(x,\vec k_T^2;P_z)$, quasi-GPD $H_1(x,\xi,t;P_z)$ and T-odd GTMDs.
In this model, two types of particles have to be considered: the pion target with mass $M$ and the quark or antiquark with mass $m$. 
A pion field $\phi$ is coupled to a quark and an antiquark using a pseudo-scalar interaction. Including isospin the interaction part of the Lagrangian reads
\begin{eqnarray} \label{eq17}
\begin{aligned}
\mathcal{L}_{\rm int}(x)=-ig_\pi \bar{\Psi}(x)\gamma_5 \vec{\tau} \cdot \vec{\phi}(x) \Psi(x),
\end{aligned}
\end{eqnarray}
where $g_\pi$ is the coupling constant and $\tau_i$ are the Pauli matrices.
In the work \cite{Ma:2019agv}, a point-like coupling have been adopted instead of a simple constant $g_\pi$ to eliminate the divergences arising after integration over large $k_T$. Furthermore the parameters of the spectator model have been determined by the authors of Ref.\cite{Ma:2019agv} through fitting the model result of unpolaried parton distribution $f_{1\pi}(x)$ to the GRV parametrization \cite{Gluck:1991ey} for the pion.
We follow the same point-like coupling form in \cite{Ma:2019agv} as
\begin{eqnarray} \label{eq18}
\begin{aligned}
g_\pi\equiv g_\pi(k_T)=g_\pi'\exp\left( -\frac{\vec k_T^2}{\bar{x}^\alpha(1-\bar{x})^\beta \lambda^2} \right)\equiv g_\pi'\exp\bigg( -\frac{\vec k_T^2}{\Lambda^2(x)} \bigg),
\end{aligned}
\end{eqnarray}
where $\bar{x}=|x|$. Here $g_\pi', \alpha, \beta$ and $\lambda$ are the model parameters.
By choosing the point-like coupling in Eq.(\ref{eq18}), the applicable range of $x$ could be $-1 < x < 1$. This kinematical region is of great interest for quasi-PDFs and quasi-GPDs.

\subsection{Reuslts for TMD $h_{1\pi}^\perp$, GPD $H_1$ and T-odd GTMDs}

We first discuss the result for TMD $h_{1\pi}^\perp$. To get nonzero results for these functions requires considering at least one-loop corrections that include effects from the Wilson line. At the leading order in $g_s^2$, one finds for the correlator in Eq.(\ref{eq1})
\begin{eqnarray}\label{eq19} 
\begin{aligned}
\Phi^{[i\sigma^{i+}\gamma_5]}=\frac{1}{2}&\int dk^- \frac{C_F g_s^2 g_\pi^2}{(2\pi)^4}\frac{i(\s{k}+m)}{k^2-m^2+i\varepsilon}\gamma_5 \frac{i(\s{P}-\s{k}-m)}{(P-k)^2-m^2+i\varepsilon}
\\
&\cdot i\int \frac{d^4 l}{(2\pi)^4} \gamma^+ \frac{-i}{l^2+i\varepsilon} \frac{1}{-l^++i\varepsilon} \frac{i(\s{k}-\s{l}+m)}{(k-l)^2-m^2+i\varepsilon} \gamma_5 \frac{i(\s{k}-\s{l}-\s{P}+m)}{(k-l-P)^2-m^2+i\varepsilon} i\sigma^{i+}\gamma_5, 
\end{aligned}
\end{eqnarray}
where $l^+$ integral is realized from taking the imaginary part of the eikonal propagator: $1/(-l^++i\varepsilon) \to -2\pi i \delta(l^+)$.
The color factor satisfies $C_F=4/3$.
Then performing the integrals for $k^-$ and $l^-$ applying contour integration together with Eq.(\ref{eq4}), one obtains
\begin{eqnarray}\label{eq20} 
\begin{aligned}
h_{1\pi}^\perp=&\frac{-mM}{2}\frac{C_F g_s^2 g_\pi^2}{(2\pi)^4}\frac{1}{\vec k_T^2(\vec k_T^2+m^2+x(x-1)M^2)}\ln\left(\frac{\vec k_T^2+m^2+M^2(x-1)x}{m^2+M^2(x-1)x}\right),
\end{aligned}
\end{eqnarray}
where we have used
\begin{eqnarray}\label{eq21} 
\begin{aligned}
\int d^2 \vec l_T \frac{\vec k_T \cdot \vec l_T}{\vec l_T^2 [(k_T-l_T)^2+m^2+M^2(x-1)x]}=-\pi\ln\left(\frac{\vec k_T^2+m^2+M^2(x-1)x}{m^2+M^2(x-1)x}\right).
\end{aligned}
\end{eqnarray}
The result in Eq.(\ref{eq20}) is the same with previous prediction in Refs.\cite{Lu:2004hu,Meissner:2008ay}. 
$h_{1\pi}^\perp$ is negative which agrees with previous expectations \cite{Burkardt:2007xm}.

The GPD $H_1$ can be extracted from the integrated quark-quark correlator in Eq.(\ref{eq8}), which reads
\begin{eqnarray}\label{eq22} 
\begin{aligned}
W^{[i\sigma^{j+}\gamma_5]}&=\int \frac{dk^-d^2 k_T}{2(2\pi)^4}\frac{g_\pi^+ g_\pi^-\text{Tr}[\gamma_5i(\s{P}-\s{k}-m)\gamma_5 i(\s{k}+\frac{1}{2}\s{\Delta}+m)i\sigma^{\alpha+}\gamma_5i(\s{k}-\frac{1}{2}\s{\Delta}+m)]}{\bigg[ (P-k)^2-m^2+i\varepsilon \bigg]\bigg[ (k-\frac{1}{2}\Delta)^2-m^2+i\varepsilon \bigg]\bigg[ (k+\frac{1}{2}\Delta)^2-m^2+i\varepsilon \bigg]},
\end{aligned}
\end{eqnarray}
where $g_\pi^\pm=g_\pi(k_T\pm \frac{1}{2}\Delta_T)$.
Performing the integrals for $k^-$ applying contour integration together with Eq.(\ref{eq9}), we obtains
\begin{equation}
H_{1}(x,\xi,\Delta_\perp) =
\begin{dcases}
0 & \qquad -1 \le x \le -\xi, \\
-\frac{g_\pi^{\prime 2} (x+\xi)(1+\xi)(1-\xi^2)}{2(2\pi)^3(1-x)}\int d^2 \vec k_\perp \frac{mM}{D_1 D_2^1}
\exp\left(-{2\vec k_\perp^2+\frac{1}{2}\vec{\Delta}_\perp^2 \over \Lambda^2(x)}\right)& \qquad -\xi \le x \le \xi, \\
-\frac{g_\pi^{\prime 2}(1-\xi^2)(1-x)}{(2\pi)^3}\int d^2\vec{k}_\perp \frac{mM}{D_1 D_2^2}\exp\left(-{2\vec k_\perp^2+\frac{1}{2}\vec\Delta_\perp^2\over\Lambda^2(x)}\right) & \qquad x \ge \xi,
\end{dcases} \label{eq23}
\end{equation}
where the $D_i$ in the denominator has the form
\begin{eqnarray}\label{eq24}
\begin{aligned}
D_1 &= (1 +\xi)^2\vec k_\perp^2+\frac{1}{4}(1-x)^2 \vec\Delta_\perp^2-(1-x)(1+\xi)\vec k_\perp \cdot \vec \Delta_\perp +(1+\xi)^2 m^2-(1-x)(x+\xi)M^2, 
\\
D_2^1 &= \xi(1-\xi^2)\vec k_\perp^2+\frac{1}{4}(1-x^2)\xi \vec\Delta_\perp^2+x(1-\xi^2) \vec{k}_\perp \cdot \vec\Delta_\perp+\xi(1-\xi^2)m^2-\xi(x^2-\xi^2)M^2,
\\
D_2^2 &= (1-\xi)^2 \vec k_\perp^2+\frac{1}{4}(1-x)^2 \vec\Delta_\perp^2+(1-x)(1-\xi)\vec{k}_\perp \cdot \vec{\Delta}_\perp+(1-\xi)^2 m^2-(1-x)(x-\xi)M^2. 
\end{aligned}
\end{eqnarray}

Then we focus on T-odd GTMDs in the spectator model. Similar to TMD case, one needs to introduce one loop diagrams for the correlator shown as 
The correlator in Eq.(\ref{eq5}) reads in the spectator model
\begin{eqnarray}\label{eq25}
\begin{aligned}
W^{[\Gamma]}=&\int dk^- \frac{C_F g_s^2 g_\pi^+ g_\pi^-}{(2\pi)^4} \cdot i\int \frac{d^4 l}{(2\pi)^4} \frac{i(\s{k}-\frac{1}{2}\s{\Delta}-\s{l}+m)}{(k-\frac{1}{2}\Delta-l)^2-m^2+i\varepsilon} \gamma_5 \frac{i(\s{k}-\s{P}-\s{l}+m)}{(k-P-l)^2-m^2+i\varepsilon} \gamma^+
\\
& \cdot \frac{-i}{l^2+i\varepsilon}  \frac{1}{-l^++i\varepsilon} \frac{i(\s{P}-\s{k}-m)}{(P-k)^2-m^2+i\varepsilon} \gamma_5 \frac{i(\s{k}+\frac{1}{2}\s{\Delta}+m)}{(k+\frac{1}{2}\Delta)^2-m^2+i\varepsilon} \Gamma.
\end{aligned}
\end{eqnarray}
Evaluating the $k^-,l^-$-integral by contour integration, the result for the four T-odd GTMDs can be cast into
\begin{equation}
H^o(x,\xi,\vec{k}_T,\vec\Delta_T) =
\begin{dcases}
0 & \quad -1 \le x \le -\xi, 
\\
0 & \quad -\xi \le x \le \xi, 
\\
\frac{8C_F g_s^2g_\pi^{\prime 2}(1-\xi^2)}{(2\pi)^5} \cdot \int d^2 \vec l_T \frac{N_H}{\vec l_T^2 D_1 D_2}\exp\left(-{2\vec k_\perp^2+\frac{1}{2}\vec \Delta_\perp^2\over \Lambda^2(x)}\right)  & \quad x \ge \xi,
\end{dcases} \label{eq26}
\end{equation}
where all the four T-odd GTMD only have nonvanishing analytical results in the DGLAP region.
The following is a compilation of the numerators of all the leading-twist T-odd GTMDs:
\begin{eqnarray}\label{eq27}
\begin{aligned}
N_{F_1^o}&=4(\xi^2-1)\vec k_T^2+4(1-\xi^2)\vec k_T \cdot \vec l_T+4\xi(x-1)\vec\Delta_T \cdot \vec k_T-2(\xi+1)(x-1)\vec\Delta_T \cdot \vec l_T+4m^2(\xi^2-1)+(x-1)^2 \vec\Delta_T^2,
\\
N_{\tilde{G}_1^o}&=-\frac{2M^2}{\vec \Delta_T^2 \vec k_T^2}\bigg( \vec\Delta_T^2(x-1)((\xi+1)\vec l_T \cdot \vec k_T -2\vec k_T^2)-2\vec k_T^2(\xi^2-1)\vec l_T \cdot \vec\Delta_T \bigg),
\\
N_{H_1^{k,o}}&=\frac{4(1-\xi^2)mM\vec l_T \cdot \vec k_T}{\vec k_T^2},
\\
N_{H_1^{\Delta,o}}&=-\frac{4mM}{\vec \Delta_T^2}\bigg[ (\xi^2-1)\vec l_T \cdot \vec\Delta_T+\vec\Delta_T^2(x-1) \bigg].
\end{aligned}
\end{eqnarray}
The denominators in Eq.(\ref{eq26}) are given by
\begin{eqnarray}\label{eq28}
\begin{aligned}
D_1&=-4\vec k_T^2(\xi-1)^2-4(\xi-1)(x-1)\vec k_T \cdot \vec\Delta_T-(x-1)^2\vec\Delta_T^2-4m^2(\xi-1)^2-4(x-1)(x-\xi)M^2,
\\
D_2&=-4(\xi+1)^2\vec k_T^2+8(\xi+1)^2\vec k_T \cdot \vec l_T-4(\xi+1)(x-1)\vec\Delta_T \cdot \vec k_T-4(\xi+1)^2 \vec l_T^2
\\&+4(\xi+1)(x-1)\vec\Delta_T \cdot \vec l_T-4m^2(\xi+1)^2-4M^2(x-1)(\xi+x)-\vec\Delta_T^2(x-1)^2.
\end{aligned}
\end{eqnarray}
If taking $|\vec \Delta_T|=0$, we work out the $\vec l_T$-integral reserving the real part of the results and obtain
\begin{equation}
H^o(x, \xi, \vec{k}_T, \vec\Delta_T) =
\begin{dcases}
0 & \quad -1 \le x \le -\xi,
\\
0 & \quad -\xi \le x \le \xi,
\\
\frac{8C_F g_s^2g_\pi^{'2}(1-\xi^2)}{(2\pi)^5 D_1}\exp\left(-{2\vec k_\perp^2 \over \Lambda^2(x)}\right)
M_H  & \quad x \ge \xi,
\end{dcases} \label{eq29}
\end{equation}
where
\begin{eqnarray}\label{eq30}
\begin{aligned}
M_{F_1^o}&=\frac{(1-\xi)\pi}{1+\xi}\ln\left( \vec k_T^2+A \over A \right),
\\
M_{\tilde{G}_1^o}&=\frac{\pi M^2(1-x)}{2(\xi+1)\vec k_T^2}\ln\left( \vec k_T^2+A \over A \right),
\\
M_{H_1^{k,o}}&=\frac{\pi m M(1-\xi)}{(1+\xi)\vec k_T^2}\ln\left( \vec k_T^2+A \over A \right),
\\
M_{H_1^{\Delta,o}}&=0,
\end{aligned}
\end{eqnarray}
with
 \begin{eqnarray}\label{eq31}
\begin{aligned}
A={ m^2(\xi+1)^2+M^2(x-1)(\xi+x) \over (\xi+1)^2 }.
\end{aligned}
\end{eqnarray}
On the other hand, when $|\vec\Delta_T| \neq 0$, using the similar method as the derivation of Eq.(\ref{eq29}), the the four T-odd GTMD results can be obtained as
\begin{equation}
H^o(x, \xi, \vec{k}_T, \vec\Delta_T) =
\begin{dcases}
0 & \quad -1 \le x \le -\xi,
\\
0 & \quad -\xi \le x \le \xi,
\\
\frac{8C_F g_s^2g_\pi^{'2}(1-\xi^2)}{(2\pi)^5 D_1}\exp\left(-{2\vec k_\perp^2+\frac{1}{2}\vec \Delta_T^2 \over \Lambda^2(x)}\right)
M_H B  & \quad x \ge \xi,
\end{dcases} \label{eq40}
\end{equation}
where
\begin{eqnarray}\label{eq41}
\begin{aligned}
M_{F_1^o}&=\pi \bigg( \frac{4|\vec k_T|}{2|\vec k_T|(\xi+1)+|\vec \Delta_T|(x-1)}-1 \bigg),
\\
M_{\tilde{G}_1^o}&=\frac{\pi M^2 (|\vec \Delta_T|(1-x)+2|\vec k_T|(\xi-1))}{2|\vec \Delta_T|\vec k_T^2(\xi+1)+|\vec k_T|\vec \Delta_T^2(x-1)},
\\
M_{H_1^{k,o}}&=\frac{2\pi|\vec k_T|mM(1-\xi)}{\vec k_T^2(2|\vec k_T|(1+\xi)+|\vec \Delta_T|(x-1))},
\\
M_{H_1^{\Delta,o}}&=\frac{2\pi|\vec \Delta_T|mM(1-\xi)}{\vec \Delta_T^2(2|\vec k_T|(1+\xi)+|\vec \Delta_T|(x-1))},
\end{aligned}
\end{eqnarray}
with
\begin{eqnarray}\label{eq42}
\begin{aligned}
B=\ln\left( \frac{4\vec k_T^2(\xi+1)^2+4\vec \Delta_T \cdot \vec k_T (\xi+1)(x-1)+4m^2(\xi+1)^2+(x-1)(4M^2(\xi+x)+2\vec \Delta_T^2(x-1))}{4m^2(\xi+1)^2+(x-1)(4M^2(\xi+x)+\vec \Delta_T^2(x-1))} \right).
\end{aligned}
\end{eqnarray}

\subsection{Results for quasi-TMD and quasi-GPD}

We start from the equal-time spatial correlation function in Eq.(\ref{eq10}), which can be written in the spectator as
\begin{eqnarray}\label{eq32}
\begin{aligned}
\Phi^{[i\sigma^{i3}\gamma_5]}=\frac{1}{2}&\int dk^0 \frac{C_F g_s^2}{(2\pi)^4}\frac{i(\s{k}+m)}{k^2-m^2+i\varepsilon}\gamma_5 \frac{i(\s{P}-\s{k}-m)}{(P-k)^2-m^2+i\varepsilon}
\\
&\cdot i\int \frac{d^4 l}{(2\pi)^4} \gamma^+ \frac{-i}{l^2+i\varepsilon} \frac{1}{-l^++i\varepsilon} \frac{i(\s{k}-\s{l}+m)}{(k-l)^2-m^2+i\varepsilon} \gamma_5 \frac{i(\s{k}-\s{l}-\s{P}+m)}{(k-l-P)^2-m^2+i\varepsilon} i\sigma^{i3}\gamma_5.
\end{aligned}
\end{eqnarray}
In order to apply contour integration we can rewrite the denominator as
\begin{eqnarray}\label{eq33}
\begin{aligned}
\frac{1}{(k^2-m^2+i\varepsilon)((P-k)^2-m^2+i\varepsilon)} \equiv \frac{1}{(k^0-k^0_-)(k^0-k^0_+)(k^0-k^{0}_{-'})(k^0-k^{0}_{+'})},
\end{aligned}
\end{eqnarray}
where $k^0_\pm, k^0_{\pm'}$ are the poles for $k^0$
\begin{eqnarray}\label{eq34}
\begin{aligned}
k^0_{\pm}&=\pm\sqrt{x^2 P_z^2+\vec k_T^2+m^2-i\varepsilon},
\\
k_{\pm'}^{0}&=P_0\pm \sqrt{(1-x)^2P_z^2+\vec k_T^2+m^2-i\varepsilon}.
\end{aligned}
\end{eqnarray}
Then according to Eq.(\ref{eq13}), the quasi-TMD reads
\begin{eqnarray}\label{eq35}
\begin{aligned}
h_{1\pi,Q(3)}^\perp(x,\vec k_T^2;P_z)&=-\frac{2\sqrt{2}mMC_F g_\pi^2 \alpha_s}{(2\pi)^4 \vec k_T^2} \cdot \ln\left(\frac{\vec k_T^2+m^2+M^2(x-1)x}{m^2+M^2(x-1)x}\right)
\\
&\cdot \bigg[ \frac{1}{(k^0_--k^0_+)(k^0_--k^0_{-'})(k^0_--k^0_{+'})}
+\frac{1}{(k^0_{-'}-k^0_-)(k^0_{-'}-k^0_+)(k^0_{-'}-k^0_{+'})} \bigg]P^+,
\end{aligned}
\end{eqnarray}
where $P^+ = 1/\sqrt{2}(\sqrt{P_z^2+M^2}+P_z)$ and $P_0=\sqrt{P_z^2+M^2}$.
It can be verified that in the limit $P_z \to \infty$, the quasi-TMD in Eq.(\ref{eq32}) reduces to the standard TMD shown as Eq.(\ref{eq20}). At the same time, $h_{1\pi,Q(0)}^\perp(x,\vec k_T^2;P_z)=-h_{1\pi,Q(3)}^\perp(x,\vec k_T^2;P_z)$.

The quasi-GPD of the pion meson can be calculated in a similar way. In the spectator model, the correlator in Eq.(\ref{eq14}) to calculate the quasi-GPD has the form:
\begin{eqnarray} \label{eq36}
\begin{aligned}
W^{[i\sigma^{\alpha3}\gamma_5]}&=-\int \frac{dk_0d^2 \vec k_T}{2(2\pi)^4}\frac{g_\pi^+ g_\pi^-\text{Tr}[(\s{P}-\s{k}+m)(\s{k}+\frac{1}{2}\s{\Delta}+m)\sigma^{\alpha3}\gamma_5(\s{k}-\frac{1}{2}\s{\Delta}+m)]}{\bigg[ (P-k)^2-m^2+i\varepsilon \bigg]\bigg[ (k-\frac{1}{2}\Delta)^2-m^2+i\varepsilon \bigg]\bigg[ (k+\frac{1}{2}\Delta)^2-m^2+i\varepsilon \bigg]}
\\
&=-\int \frac{dk_0d^2 \vec k_T}{2(2\pi)^4}\frac{g_\pi^+ g_\pi^-\text{Tr}[(\s{P}-\s{k}+m)(\s{k}+\frac{1}{2}\s{\Delta}+m)\sigma^{\alpha3}\gamma_5(\s{k}-\frac{1}{2}\s{\Delta}+m)]}{(k^0-k^0_{1-})(k^0-k^0_{1+})(k^0-k^0_{2-})(k^0-k^0_{2+})(k^0-k^0_{3-})(k^0-k^0_{3+})}.
\end{aligned}
\end{eqnarray}
After performing the $k^0$-integral using the contour integration, we write down the analytical result of the quasi-GPD
as follows according to Eq.(\ref{eq16})
\begin{eqnarray} \label{eq37}
\begin{aligned}
H_{1,Q(3)}(x,\xi,\Delta_T;P_z)&=\int d^2 \vec k_T \frac{4g_\pi^+ g_\pi^- mM\delta P_z}{2(2\pi)^3}\sum_{i=0}^{3}\frac{1}{D_i},
\end{aligned}
\end{eqnarray}
where
\begin{eqnarray} \label{eq38}
\begin{aligned}
&D_1=(k^0_{1-}-k^0_{1+})(k^0_{1-}-k^0_{2+})(k^0_{1-}-k^0_{2-})(k^0_{1-}-k^0_{3+})(k^0_{1-}-k^0_{3-}),
\\
&D_2=(k^0_{2-}-k^0_{1+})(k^0_{2-}-k^0_{1-})(k^0_{2-}-k^0_{2+})(k^0_{2-}-k^0_{3+})(k^0_{2-}-k^0_{3-}),
\\
&D_3=(k^0_{3-}-k^0_{1+})(k^0_{3-}-k^0_{1-})(k^0_{3-}-k^0_{2+})(k^0_{3-}-k^0_{2-})(k^0_{3-}-k^0_{3+}).
\end{aligned}
\end{eqnarray}
The poles coming from the denominator are given by
\begin{eqnarray} \label{eq39}
\begin{aligned}
&k^0_{1\pm}=\delta P_z \pm \sqrt{(1-x)^2P_z^2+\vec k_T^2+m^2-i\varepsilon},
\\
&k^0_{2\pm}=-\xi P_z \pm \sqrt{(x+\delta \xi)^2 P_z^2+\left( k_T-\frac{\Delta_T}{2} \right)^2+m^2-i\varepsilon},
\\
&k^0_{3\pm}=\xi P_z \pm \sqrt{(x-\delta \xi)^2 P_z^2+\left( k_T+\frac{\Delta_T}{2} \right)^2+m^2-i\varepsilon},
\end{aligned}
\end{eqnarray}
with $\delta=P_0/P_z=1/P_z\sqrt{-t/4+P_z^2+M^2}$. Similarly, $H_{1,Q(0)}(x,\xi,\Delta_T;P_z)=\frac{1}{\delta}H_{1,Q(3)}(x,\xi,\Delta_T;P_z)$.

\section{Numerical calculation} 
\label{IV}

In order to fix the parameters of the spectator model, 
the authors of Ref.\cite{Ma:2019agv} fit the model result of unpolaried 
parton distribution $f_{1\pi}(x)$ to the GRV parametrization \cite{Gluck:1991ey} for the pion.
We adopt the fitted values for the parameters $g_\pi'=6.316, \lambda=0.855, \alpha=0$ and $\beta=1$.
We make a preliminary estimate for choosing the strong coupling $\alpha_s \approx 0.3$ and adopting the quark mass $m=0.3$GeV.

\begin{figure}[htp]
\centering
\includegraphics[height=3.5cm,width=5cm]{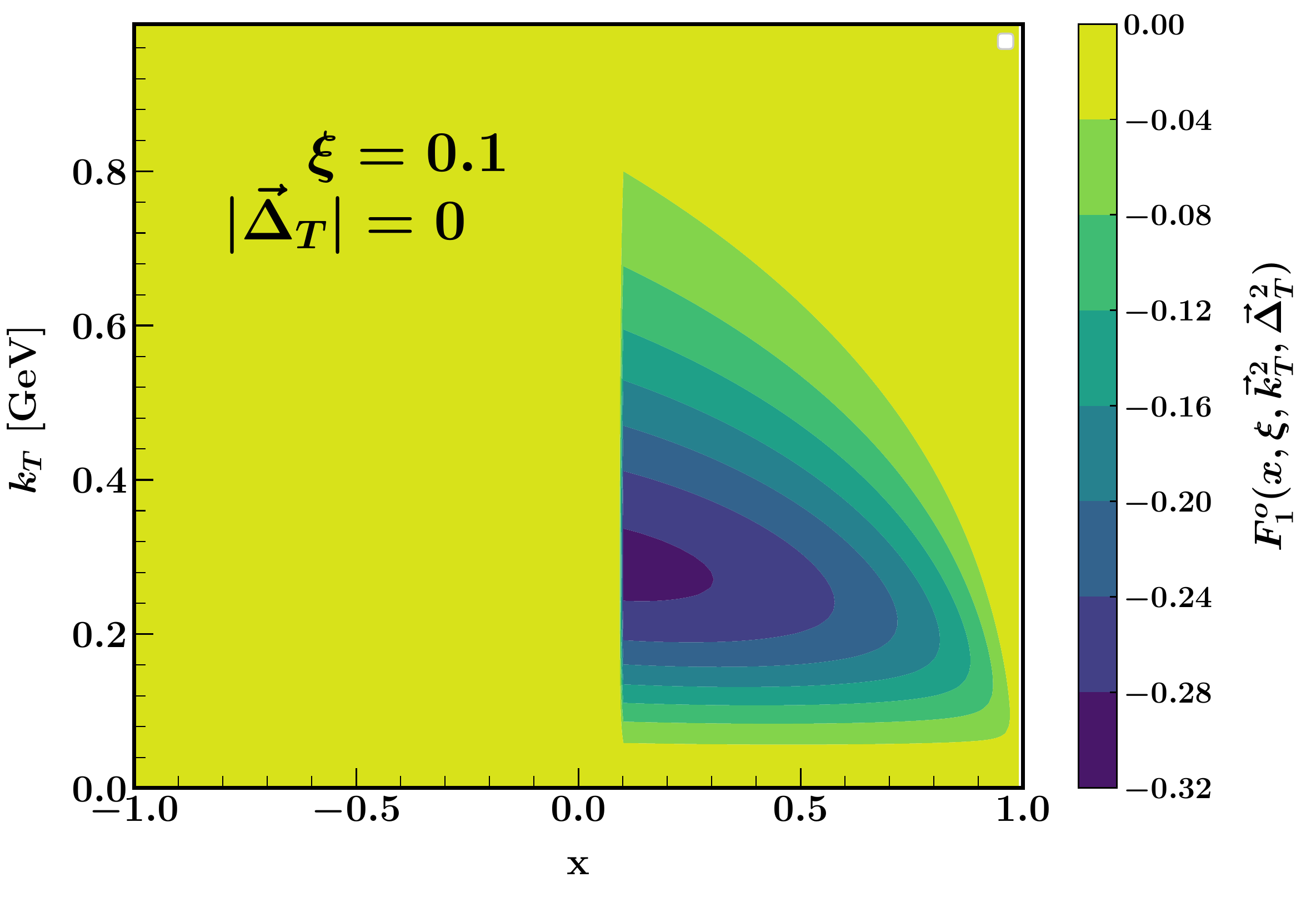}
\includegraphics[height=3.5cm,width=5cm]{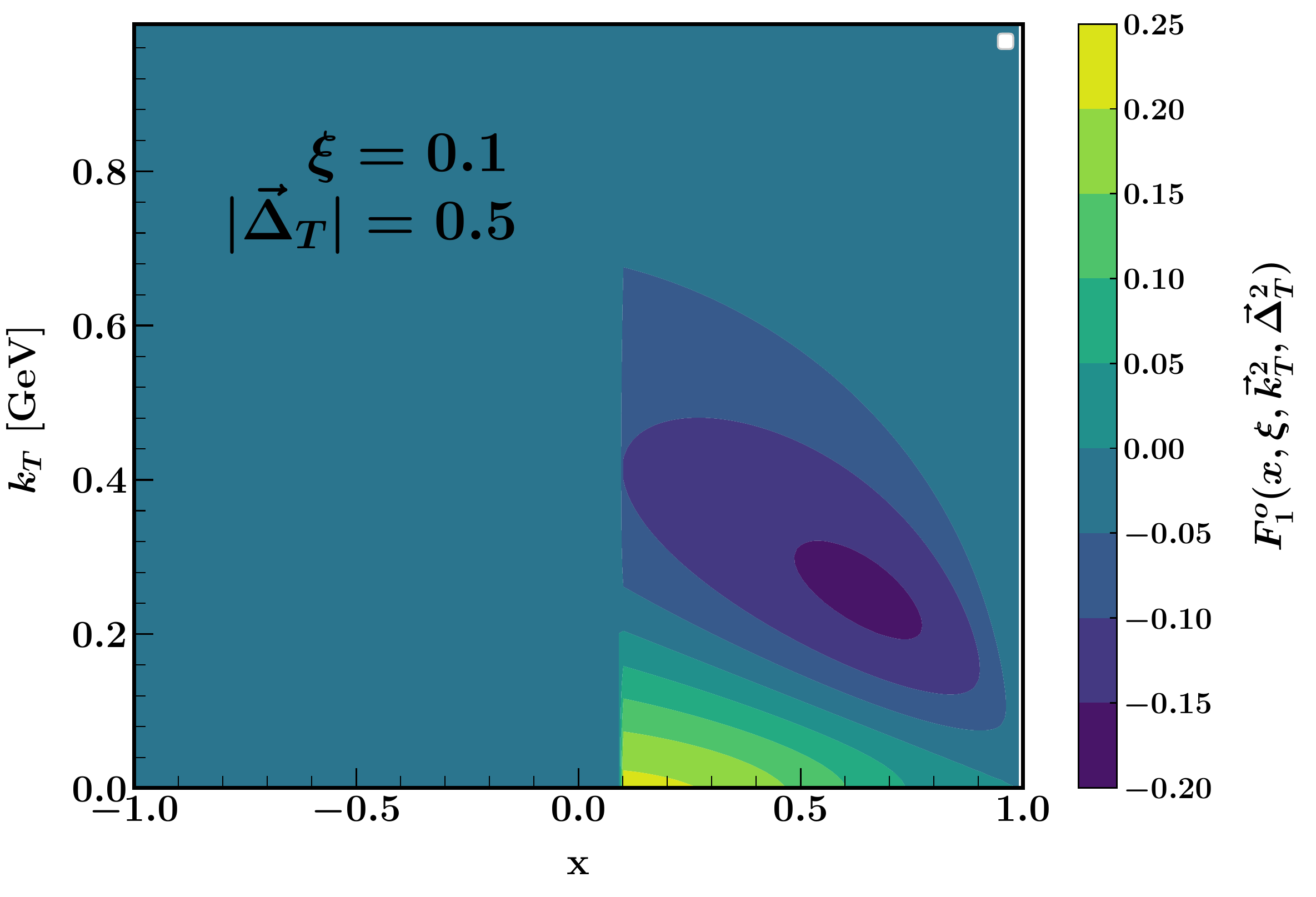}\\
\includegraphics[height=3.5cm,width=5cm]{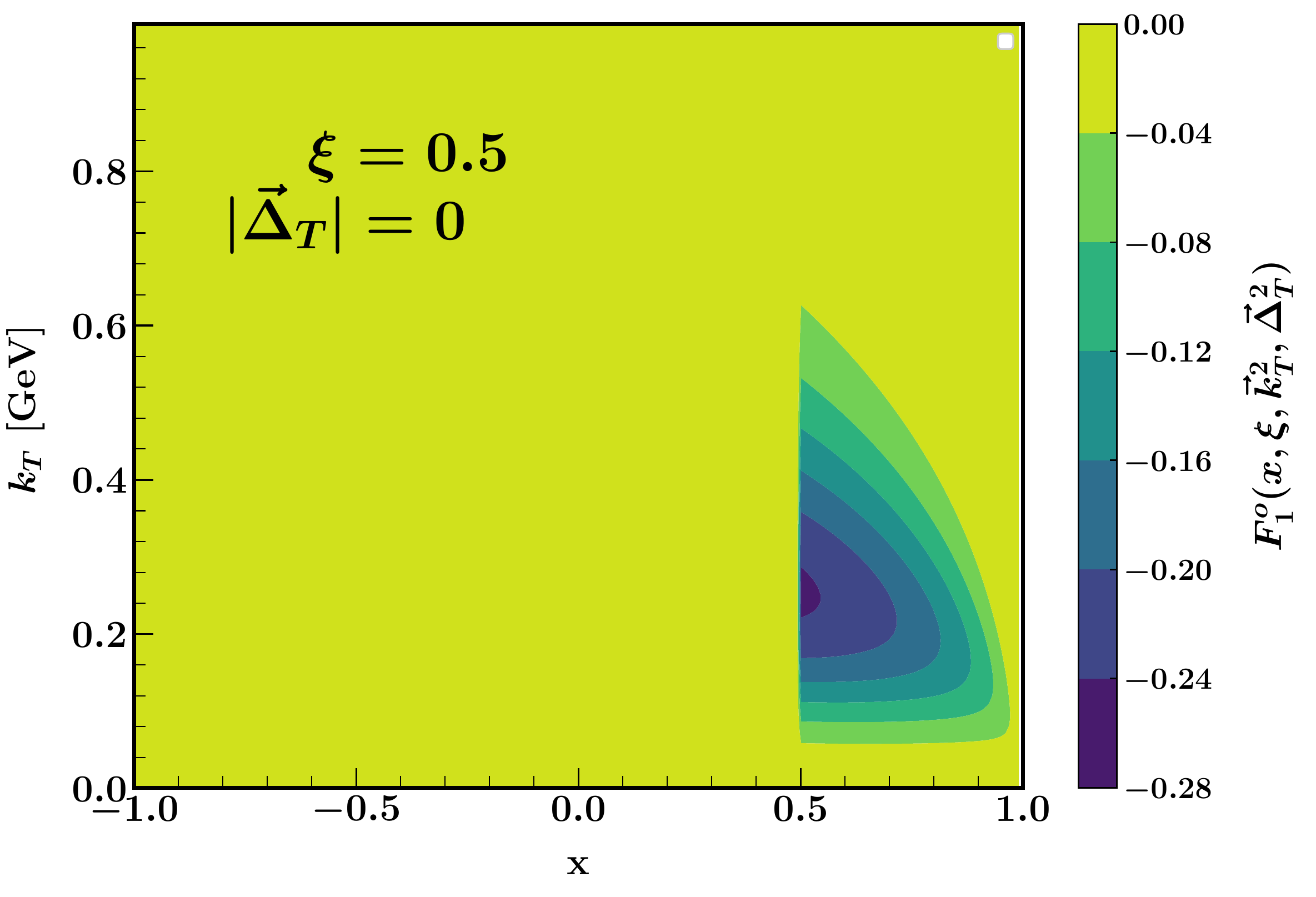}
\includegraphics[height=3.5cm,width=5cm]{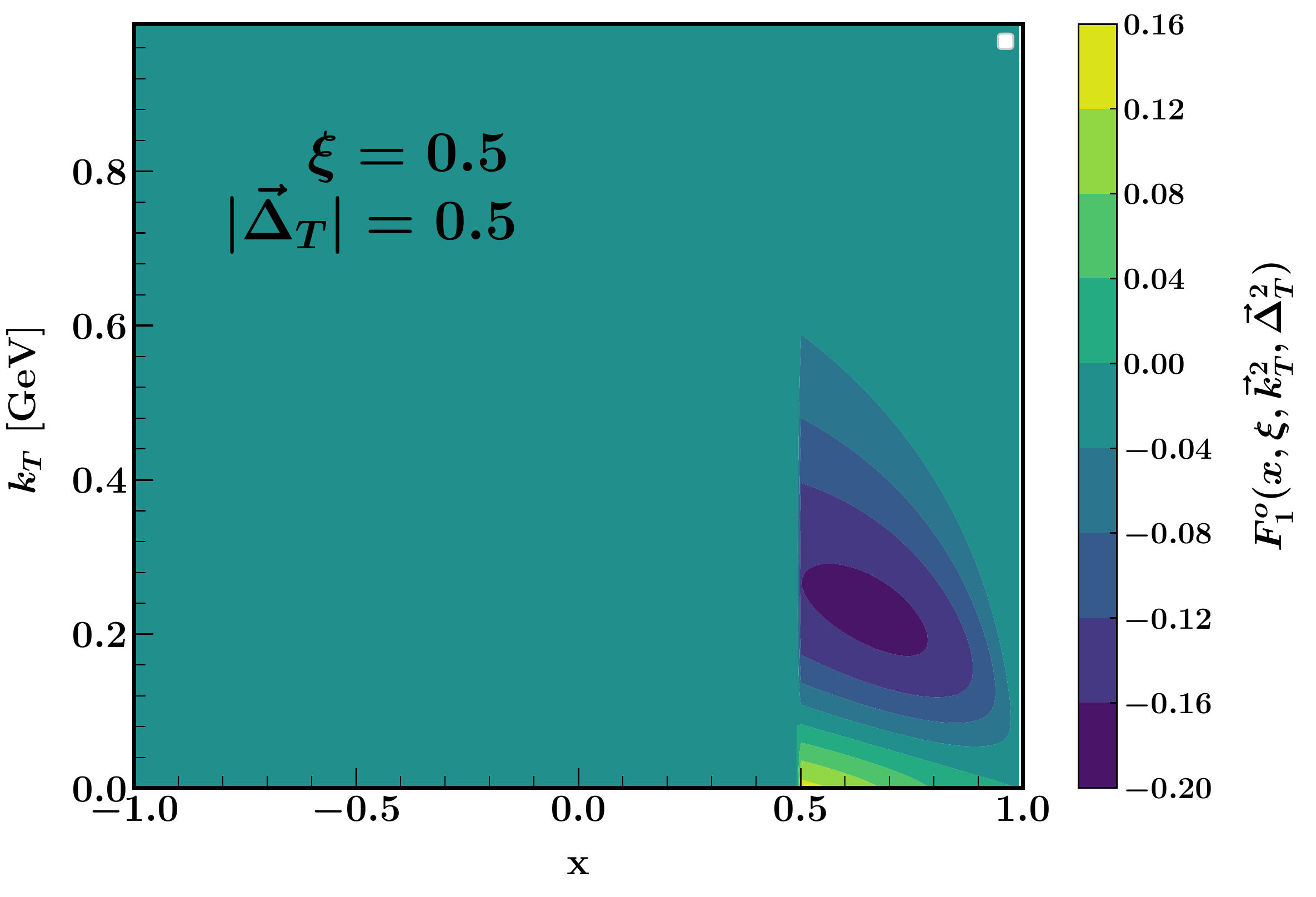}
\caption{\normalsize The T-odd GTMD $F_1^o(x,\xi,\vec k_T^2,\vec\Delta_T^2)$ as functions of $x$ and $|\vec k_T|$ for different $\xi$ and $|\vec\Delta_T|$ values. 
The upper and lower panels display $F_1^o(x,\xi,\vec k_T^2,\vec\Delta_T^2)$ at $\xi=0.1$ and $0.5$, respectively. 
The left and right panels $F_1^o(x,\xi,\vec k_T^2,\vec\Delta_T^2)$ at $|\vec\Delta_T|=0$ and $0.5$, respectively.}
\label{fig1}
\end{figure}
\begin{figure}[htp]
\centering
\includegraphics[height=3.5cm,width=5cm]{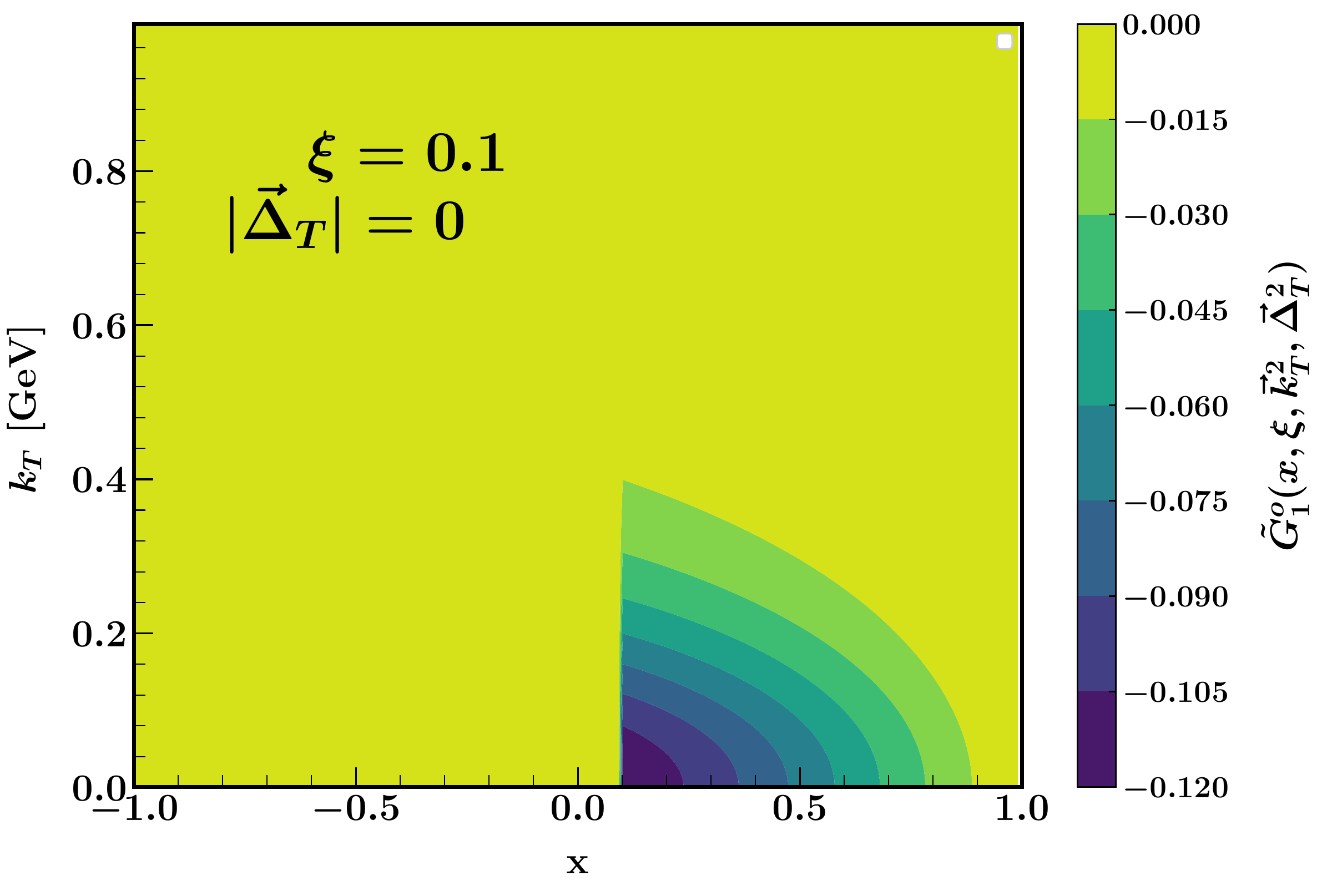}
\includegraphics[height=3.5cm,width=5cm]{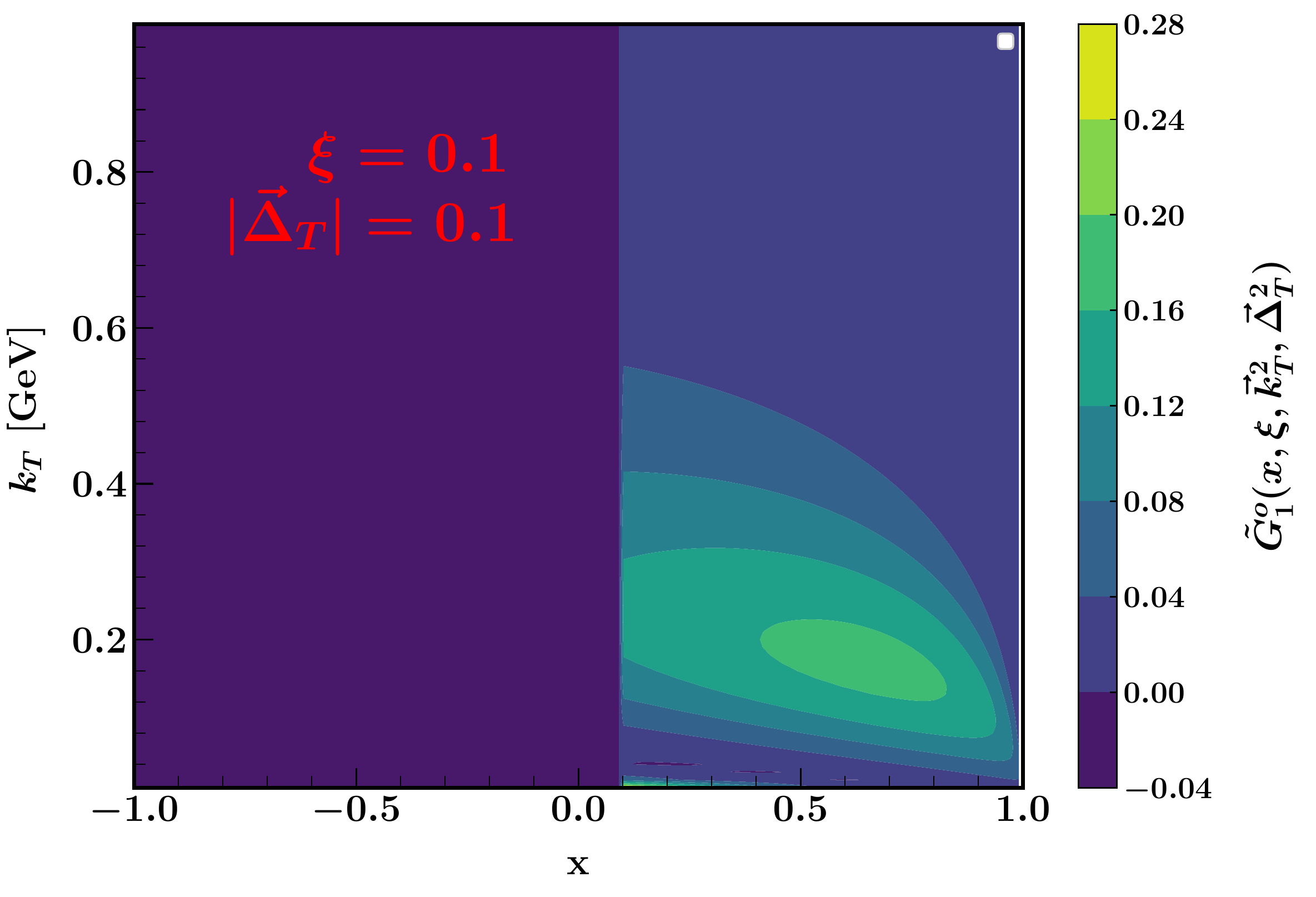}\\
\includegraphics[height=3.5cm,width=5cm]{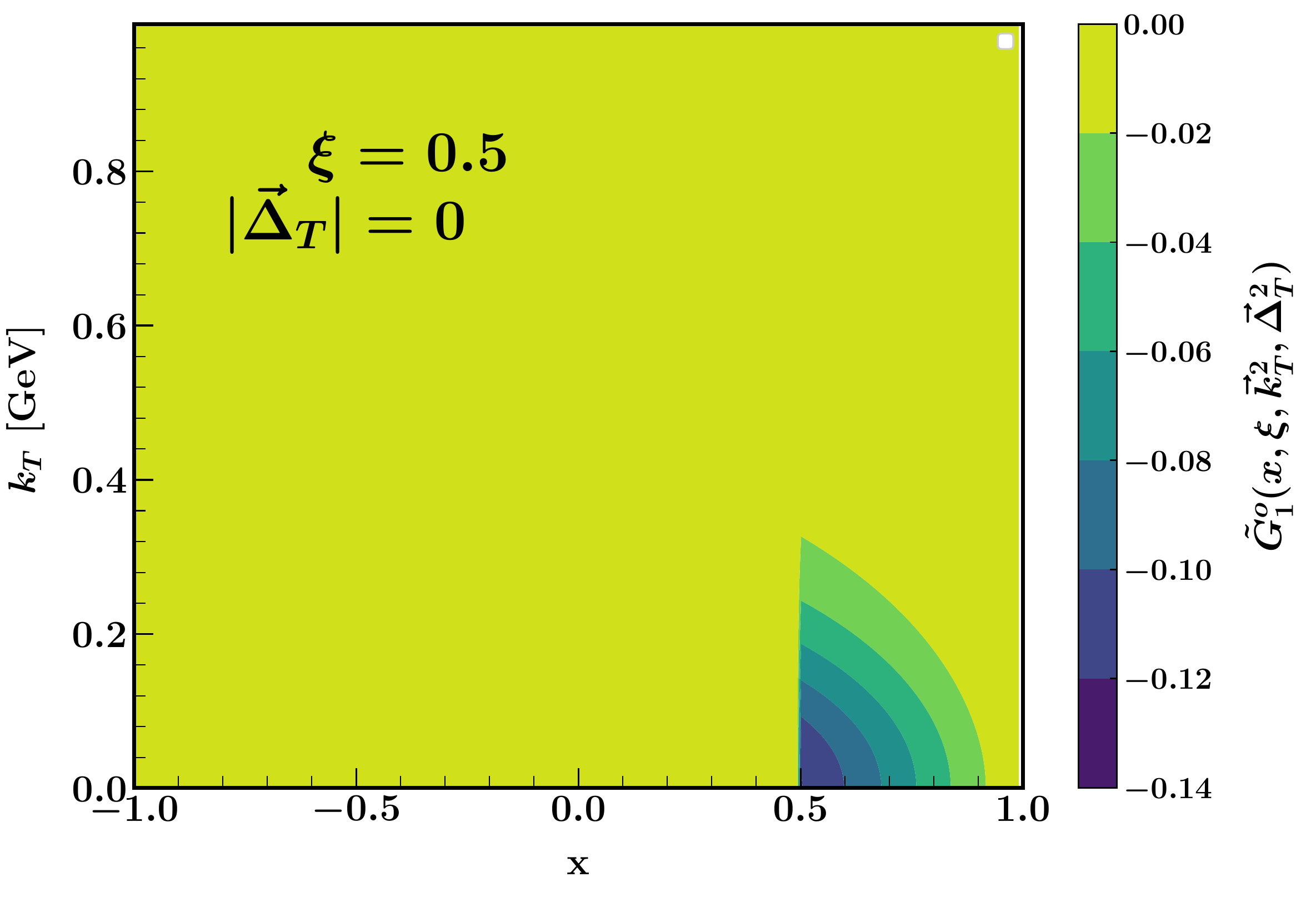}
\includegraphics[height=3.5cm,width=5cm]{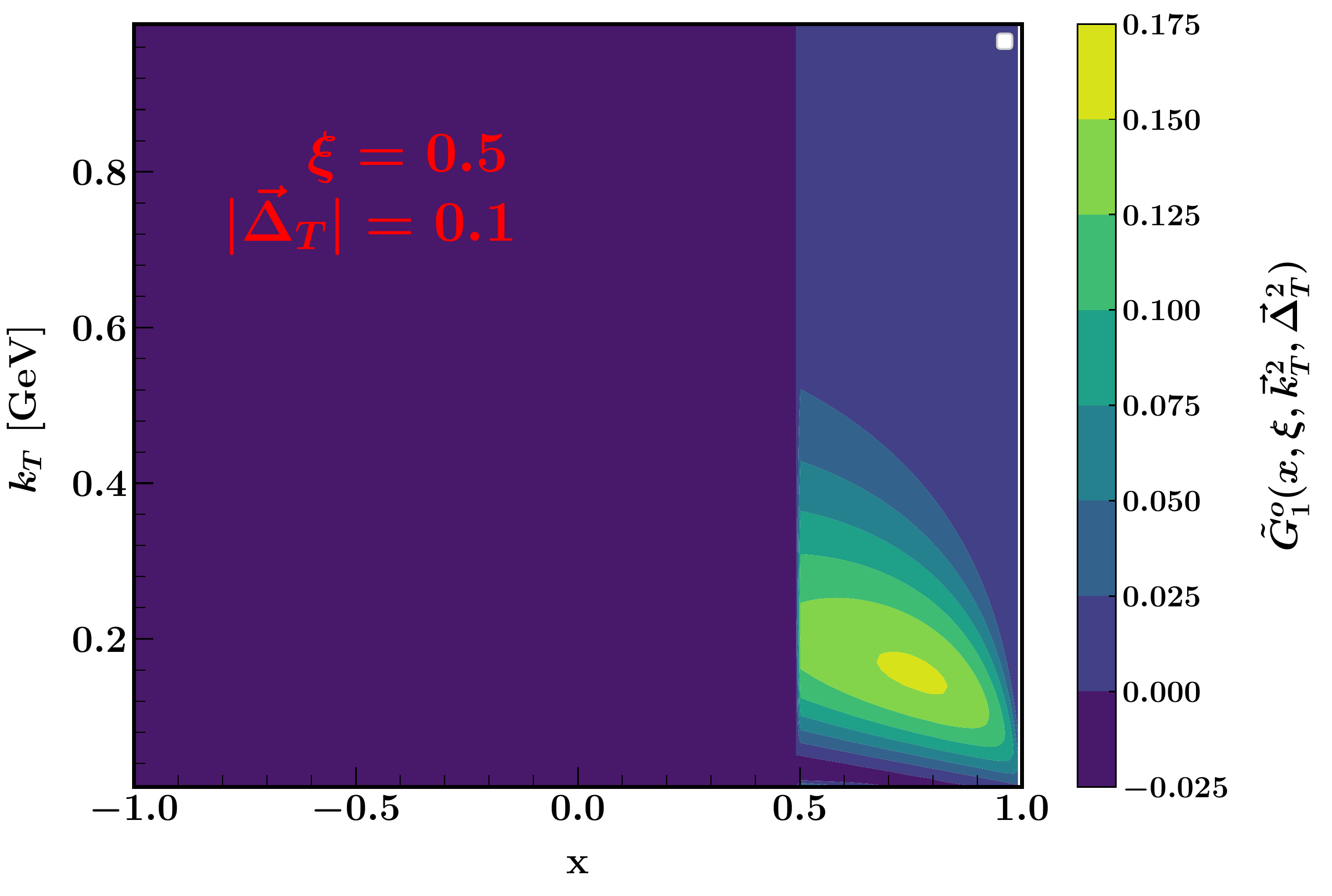}
\caption{\normalsize The T-odd GTMD $\tilde{G}_1^o(x,\xi,\vec k_T^2,\vec\Delta_T^2)$ as functions of $x$ and $|\vec k_T|$ for different $\xi$ and $|\vec\Delta_T|$ values. 
The upper and lower panels display $\tilde{G}_1^o(x,\xi,\vec k_T^2,\vec\Delta_T^2)$ at $\xi=0.1$ and $0.5$, respectively. 
The left and right panels $\tilde{G}_1^o(x,\xi,\vec k_T^2,\vec\Delta_T^2)$ at $|\vec\Delta_T|=0$ and $0.1$, respectively.}
\label{fig2}
\end{figure}
\begin{figure}[htp]
\centering
\includegraphics[height=3.5cm,width=5cm]{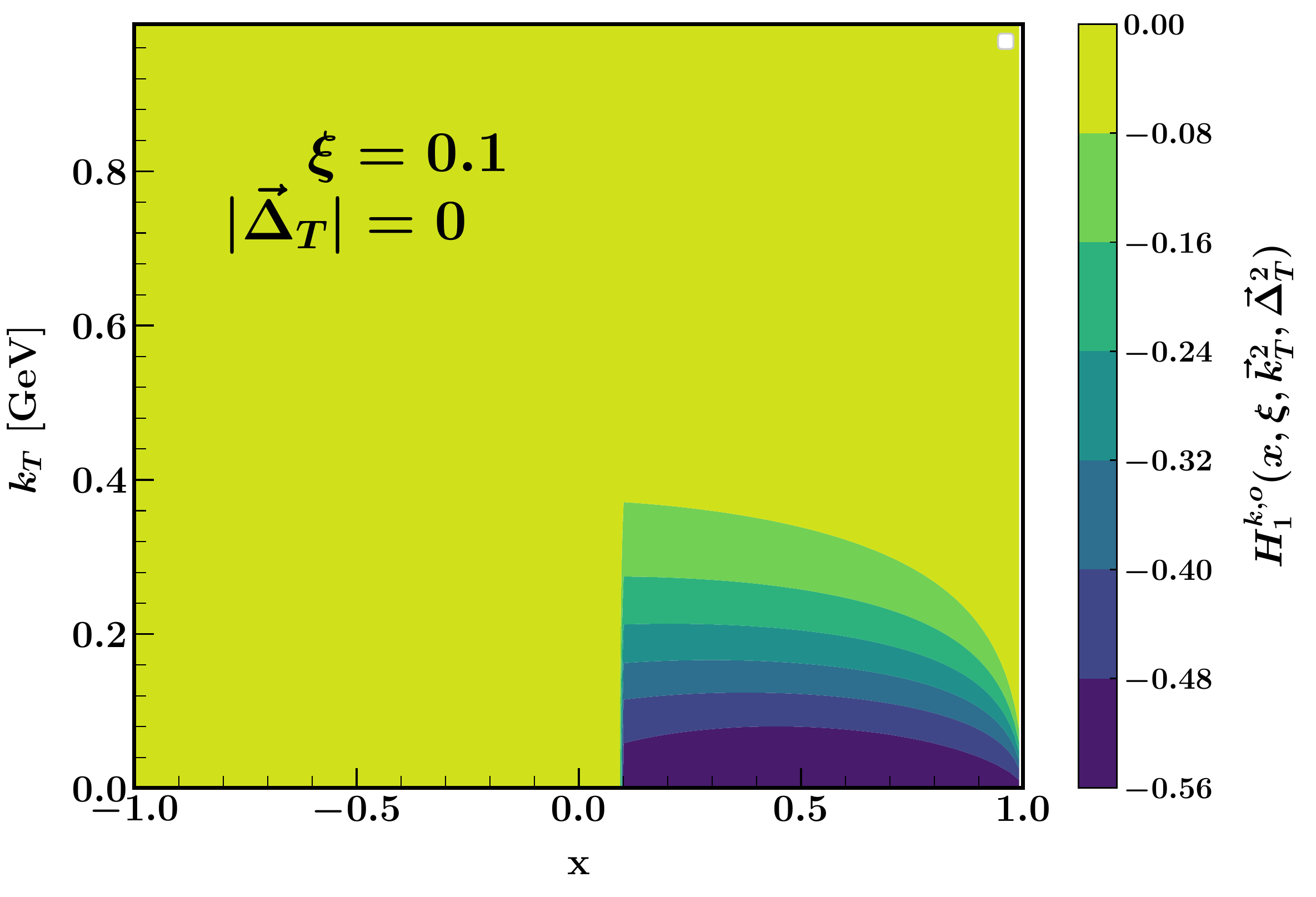}
\includegraphics[height=3.5cm,width=5cm]{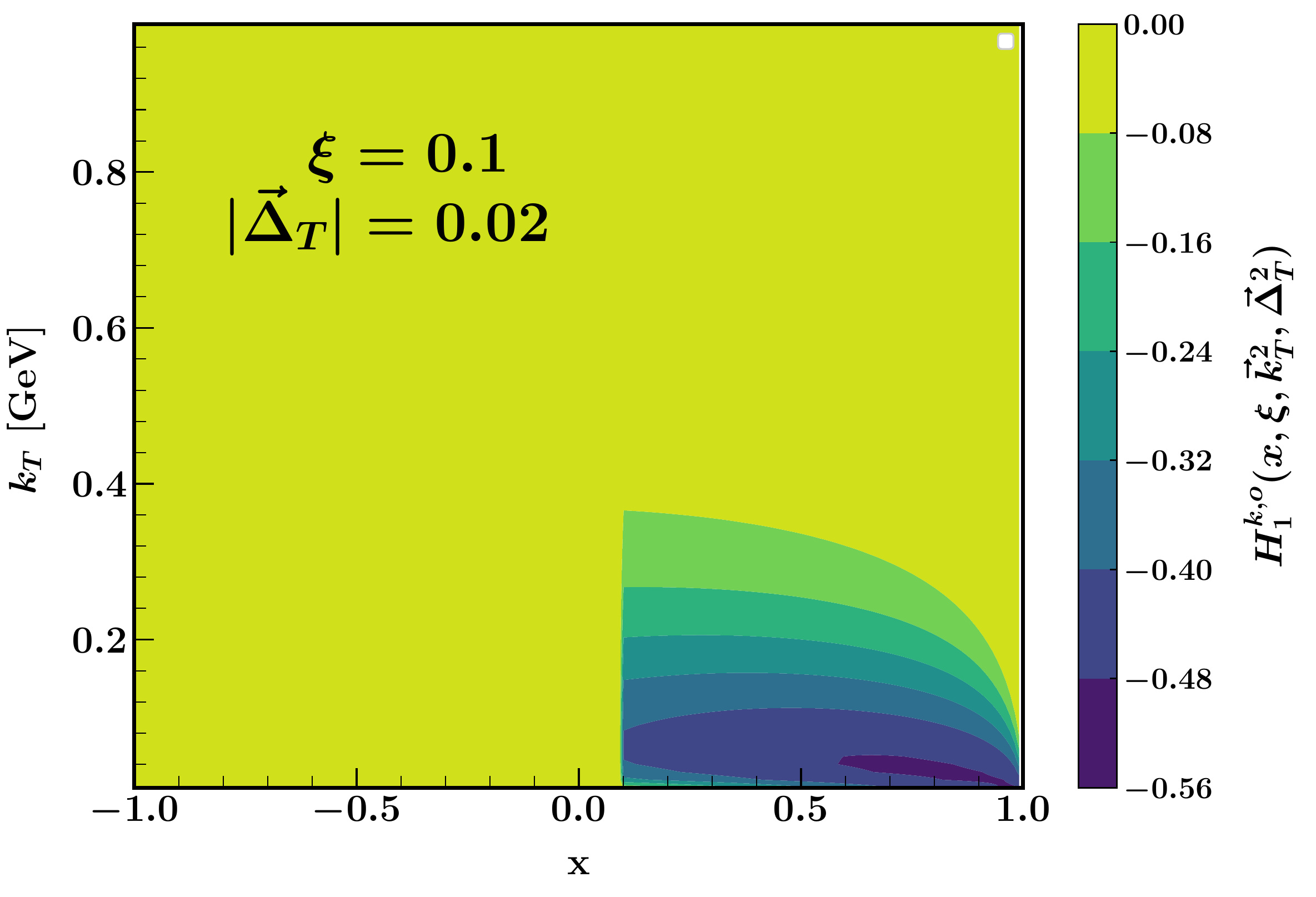}\\
\includegraphics[height=3.5cm,width=5cm]{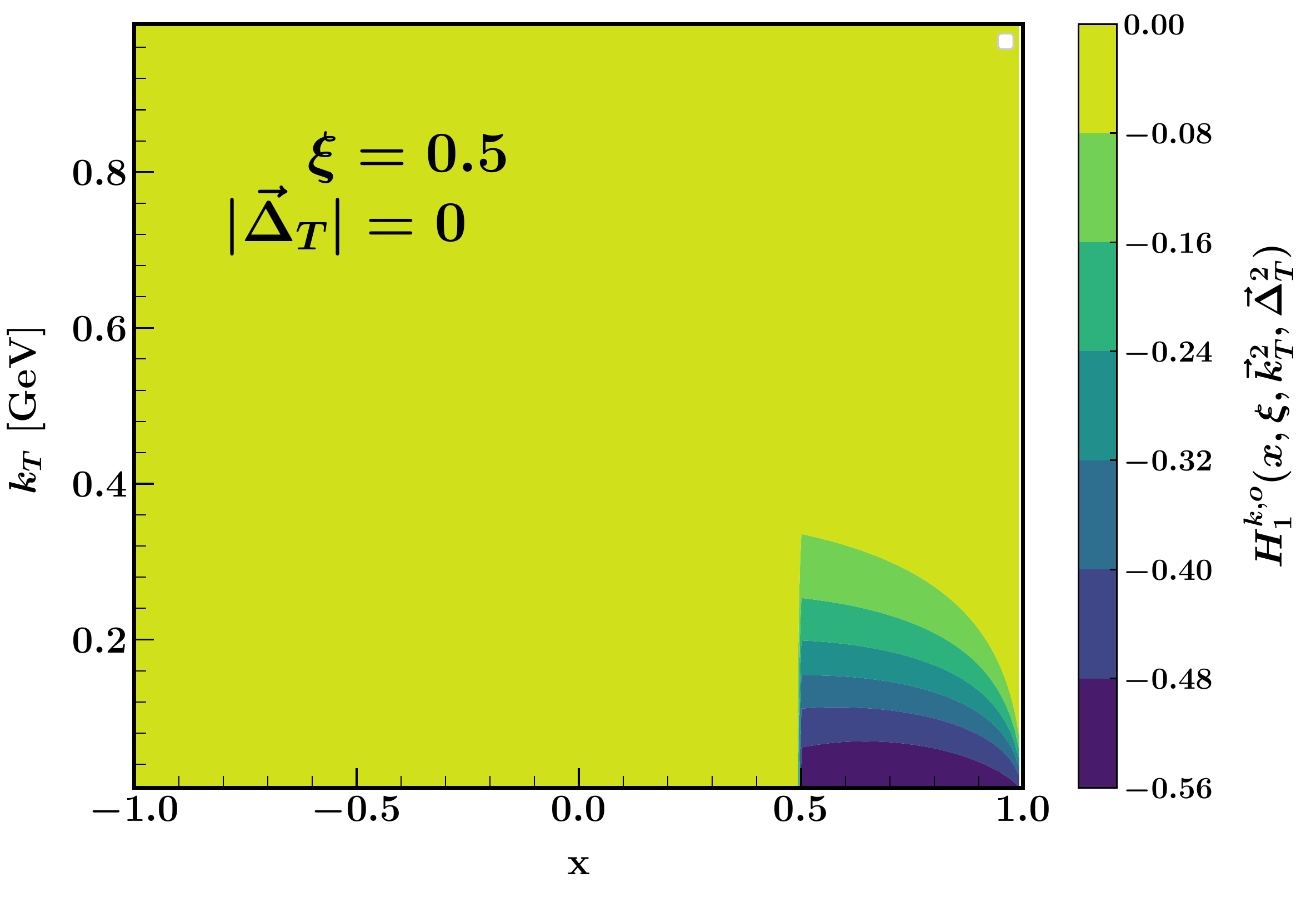}
\includegraphics[height=3.5cm,width=5cm]{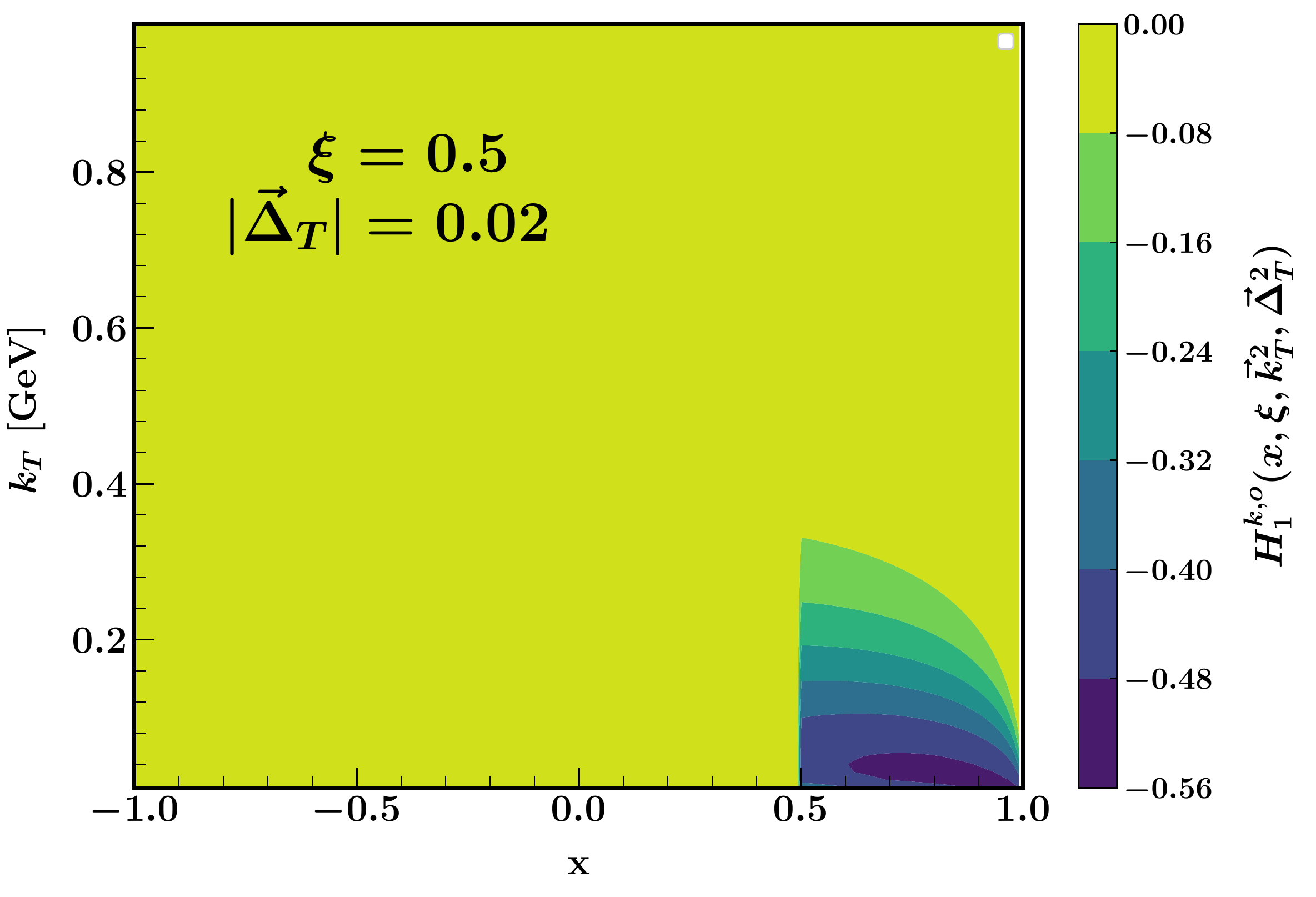}
\caption{\normalsize The T-odd GTMD $H_1^{k,o}(x,\xi,\vec k_T^2,\vec\Delta_T^2)$ as functions of $x$ and $|\vec k_T|$ for different $\xi$ and $|\vec\Delta_T|$ values. 
The upper and lower panels display $H_1^{k,o}(x,\xi,\vec k_T^2,\vec\Delta_T^2)$ at $\xi=0.1$ and $0.5$, respectively. 
The left and right panels $H_1^{k,o}(x,\xi,\vec k_T^2,\vec\Delta_T^2)$ at $|\vec\Delta_T|=0$ and $0.02$, respectively.}
\label{fig3}
\end{figure}
\begin{figure}[htp]
\centering
\includegraphics[height=3.5cm,width=5cm]{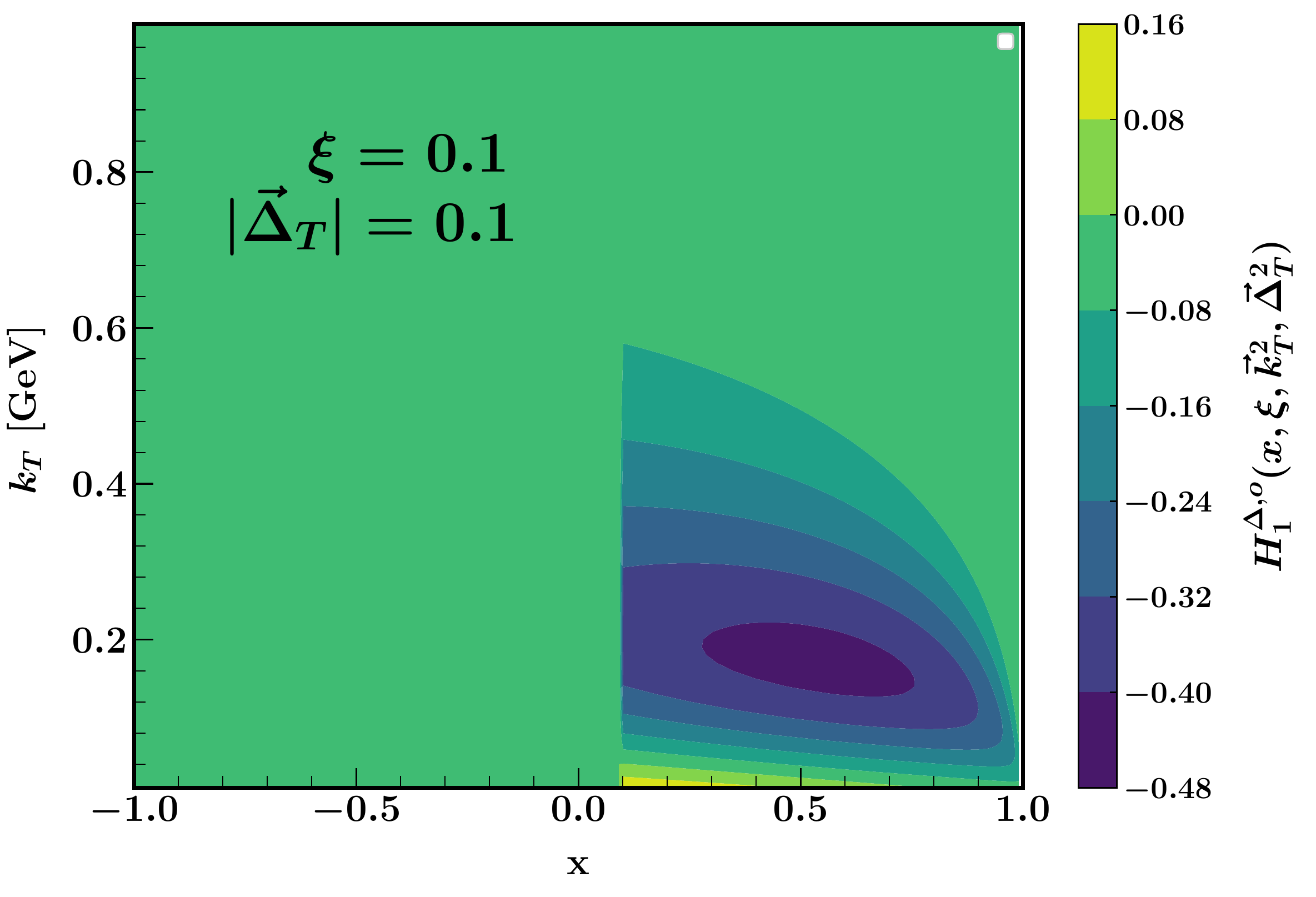}
\includegraphics[height=3.5cm,width=5cm]{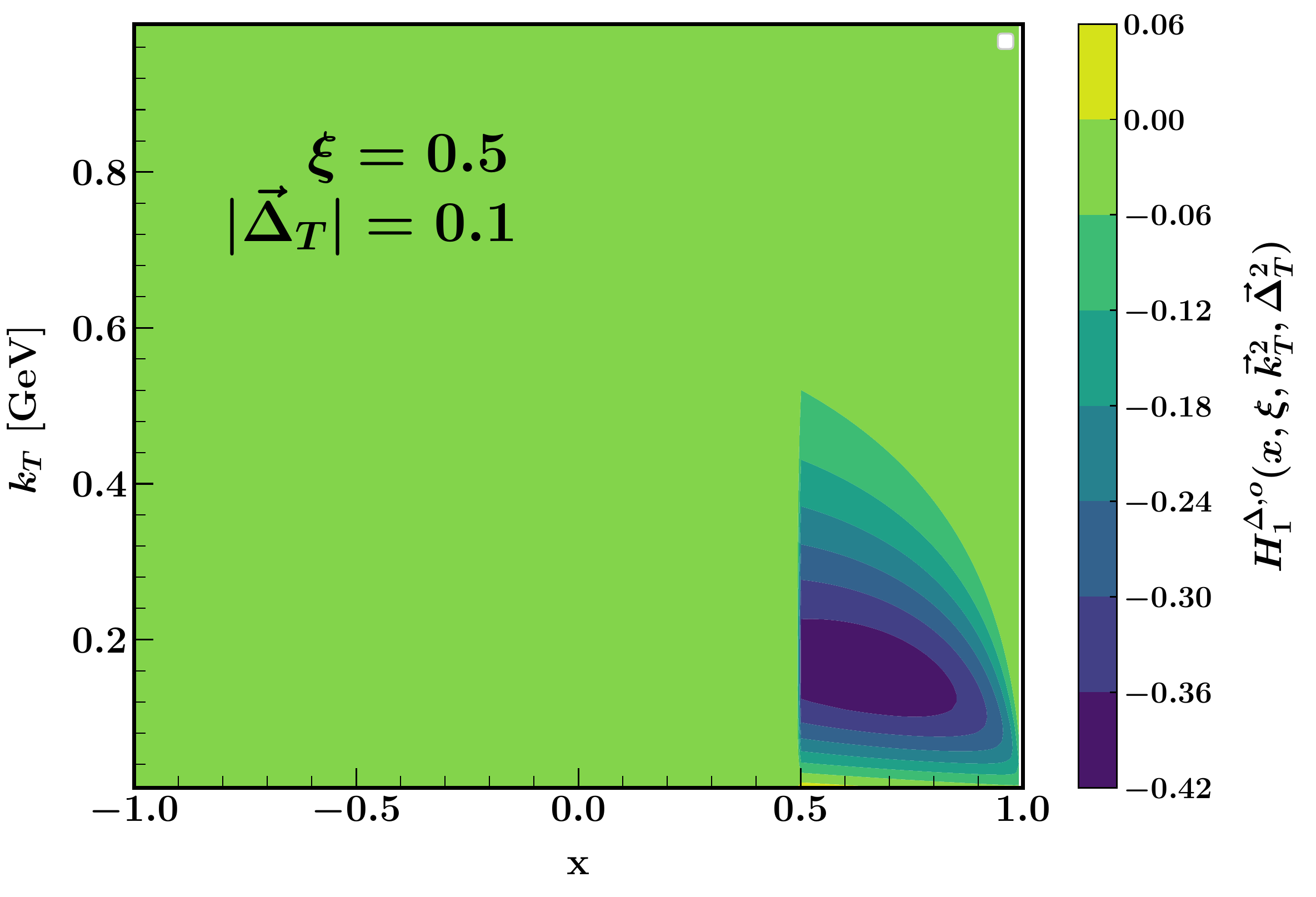}
\caption{\normalsize The T-odd GTMD $H_1^{\Delta,o}(x,\xi,\vec k_T^2,\vec\Delta_T^2)$ as functions of $x$ and $|\vec k_T|$ for different $\xi$ values. 
The left and right panels $H_1^{\Delta,o}(x,\xi,\vec k_T^2,\vec\Delta_T^2)$ at $|\vec\Delta_T|=0.1$ with $\xi=0.1$ and $0.5$, respectively.}
\label{fig4}
\end{figure}

Firstly, we depict the four T-odd GTMD results considering different values of $\xi$ and $\vec\Delta_T$ in Figs.\ref{fig1}-\ref{fig4}.
The T-odd GTMD $F_1^o(x,\xi,\vec k_T^2,\vec\Delta_T^2)$ as functions of $x$ and $\vec k_T$ are shown in Fig.\ref{fig1}. 
According to Eq.(\ref{eq26}), only in $x>\xi$ region, the T-odd GTMD has nonvanishing value. 
In the upper, left panel of Fig.\ref{fig1}, we plot the T-odd GTMD $F_1^o(x,\xi,\vec k_T^2,\vec\Delta_T^2)$ at $\xi=0.1$ and $|\vec\Delta_T|=0$. 
Note that in this case the result of $F_1^o(x,\xi,\vec k_T^2,\vec\Delta_T^2)$ is negative. 
One can also find that as the $x$ value becomes larger, the maximum value of $k_T$ resulting in 
nonvanishing $F_1^o(x,\xi,\vec k_T^2,\vec\Delta_T^2)$ becomes smaller and at $x=0.1$, the maximum value of $k_T$ is about 0.8GeV.
For a fixed $x$ value, as the value of $k_T$ increases, the $F_1^o(x,\xi,\vec k_T^2,\vec\Delta_T^2)$ result becomes smaller first and then larger. 
At a fixed value $x=0.1$, the $F_1^o(x,\xi,\vec k_T^2,\vec\Delta_T^2)$ reachs the minimum when $k_T$ is about 0.3. 
We depict the $F_1^o(x,\xi,\vec k_T^2,\vec\Delta_T^2)$ at $\xi=0.1$ and $|\vec\Delta_T|=0.5$ in the upper, right panel of Fig.\ref{fig1}. 
Roughly speaking, it can be seen that the $x$-$k_T$ regions below straight line $y=-0.25x+0.2$ are related to positive $F_1^o(x,\xi,\vec k_T^2,\vec\Delta_T^2)$.
Then we compare the two upper panels in Fig.\ref{fig1} and find that the $F_1^o(x,\xi,\vec k_T^2,\vec\Delta_T^2)$ result in the right panel 
is larger than that in the left panel at the same $x$-$k_T$ point.
Comparing the upper panels with the lower panels, we emphasize that the two corresponding contours at $x > 0.5$ have nearly the same shape.

Fig.\ref{fig2} depicts the T-odd GTMD $\tilde{G}_1^o(x,\xi,\vec k_T^2,\vec\Delta_T^2)$ as functions of $x$ and $\vec k_T$, 
where the paremeter values of contours are the same as Fig.\ref{fig1}. 
In two left panels, the $\tilde{G}_1^o(x,\xi,\vec k_T^2,\vec\Delta_T^2)$ result increases as $x$ or $k_T$ increases. 
Comparing two left panels, we find that the $\tilde{G}_1^o(x,\xi,\vec k_T^2,\vec\Delta_T^2)$ result in the lower panel 
is slightly smaller than that in the upper panel at the same $x$-$k_T$ point; 
While the $\tilde{G}_1^o(x,\xi,\vec k_T^2,\vec\Delta_T^2)$ results in two right panels are usually positive. 
Unlike two right panels in Fig.\ref{fig1} where we plot contours at $|\vec\Delta_T|=0.5$, we show the result at $|\vec\Delta_T|=0.1$. 
Such difference indicates that the allowable maximum of $|\vec\Delta_T|$ 
for reaching nonvanishing T-odd GTMD $\tilde{G}_1^o(x,\xi,\vec k_T^2,\vec\Delta_T^2)$ becomes smaller in terms of that for $F_1^o(x,\xi,\vec k_T^2,\vec\Delta_T^2)$.
We plot the T-odd GTMD $H_1^{k,o}(x,\xi,\vec k_T^2,\vec\Delta_T^2)$ as functions of $x$ and $\vec k_T$ in Fig.\ref{fig3}, 
where the paremeter values of contours are the same as Fig.\ref{fig1}. All four panels basically show the same shape of contours.
Moreover, the maximum absolute value of the T-odd GTMD $H_1^{k,o}(x,\xi,\vec k_T^2,\vec\Delta_T^2)$ can achieve 0.56 which is a larger value than those in Figs.\ref{fig1}-\ref{fig2}.
In Fig.\ref{fig4}, the T-odd GTMD $H_1^{\Delta,o}(x,\xi,\vec k_T^2,\vec\Delta_T^2)$ as a function of $x$ and $\vec k_T$
at $|\vec\Delta_T|=0.1$ has been shown. Note that this GTMD becomes zero when $|\vec\Delta_T|=0$.

\begin{figure}[htp]
\centering
\includegraphics[height=3.5cm,width=5cm]{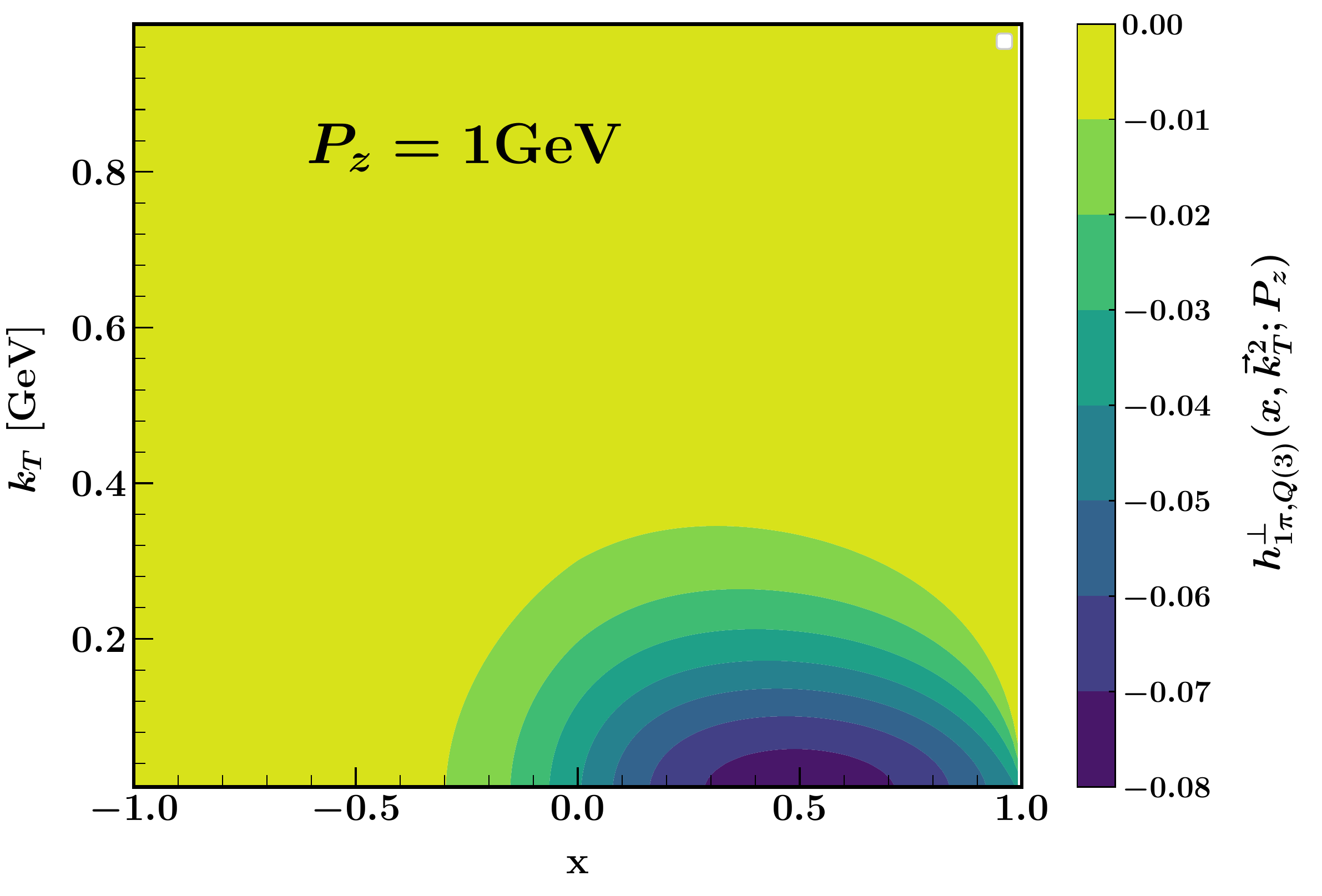}
\includegraphics[height=3.5cm,width=5cm]{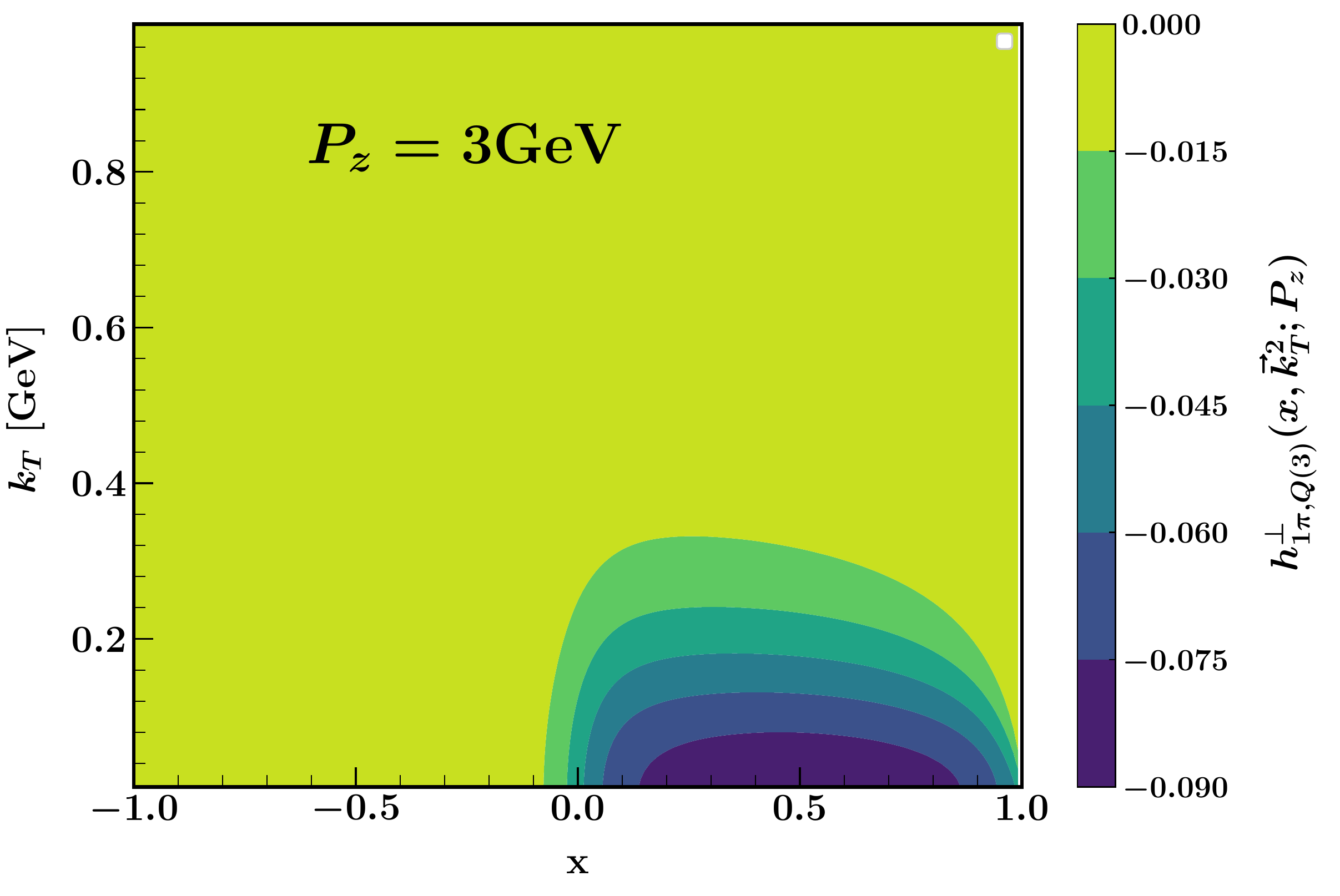}\\
\includegraphics[height=3.5cm,width=5cm]{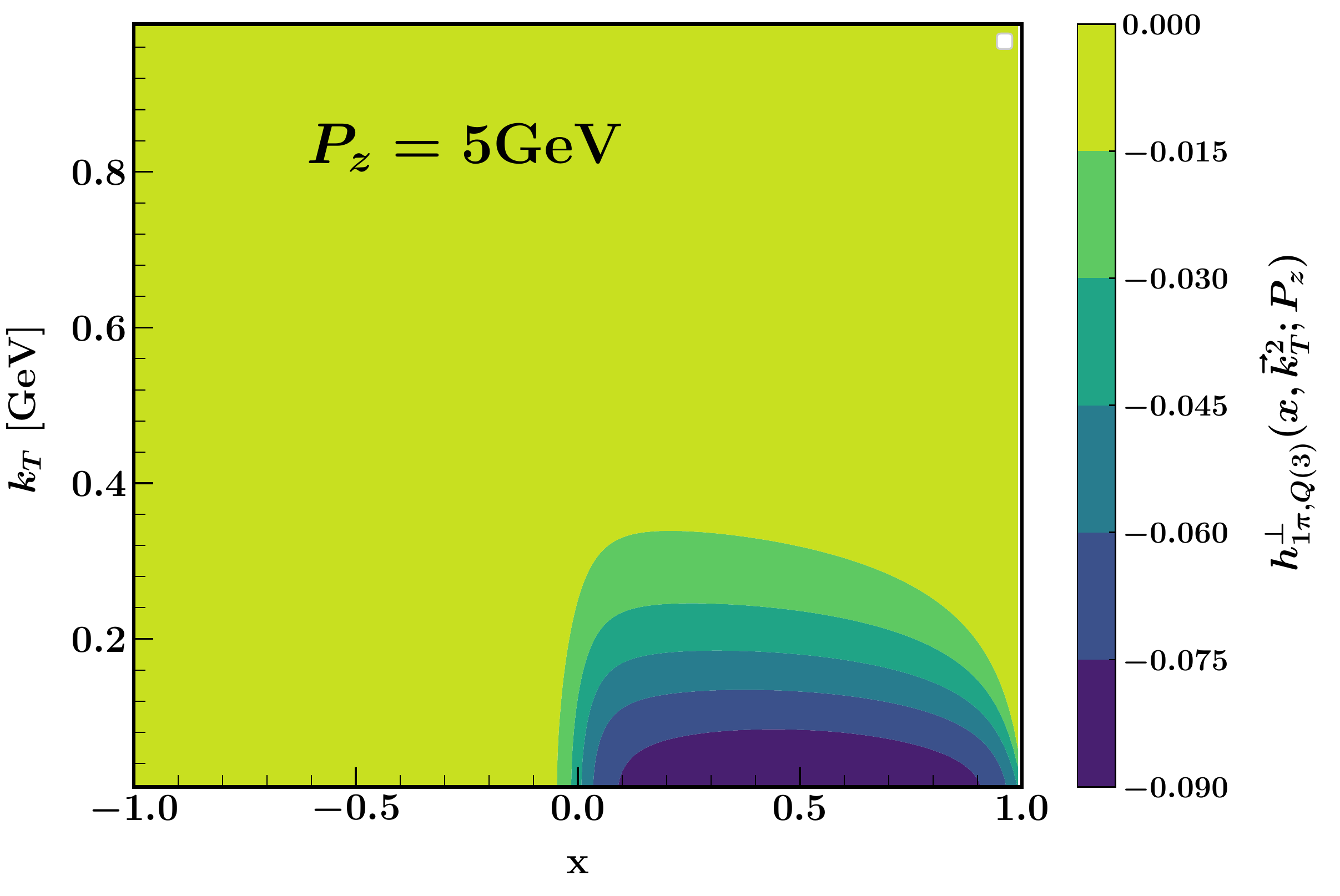}
\includegraphics[height=3.5cm,width=5cm]{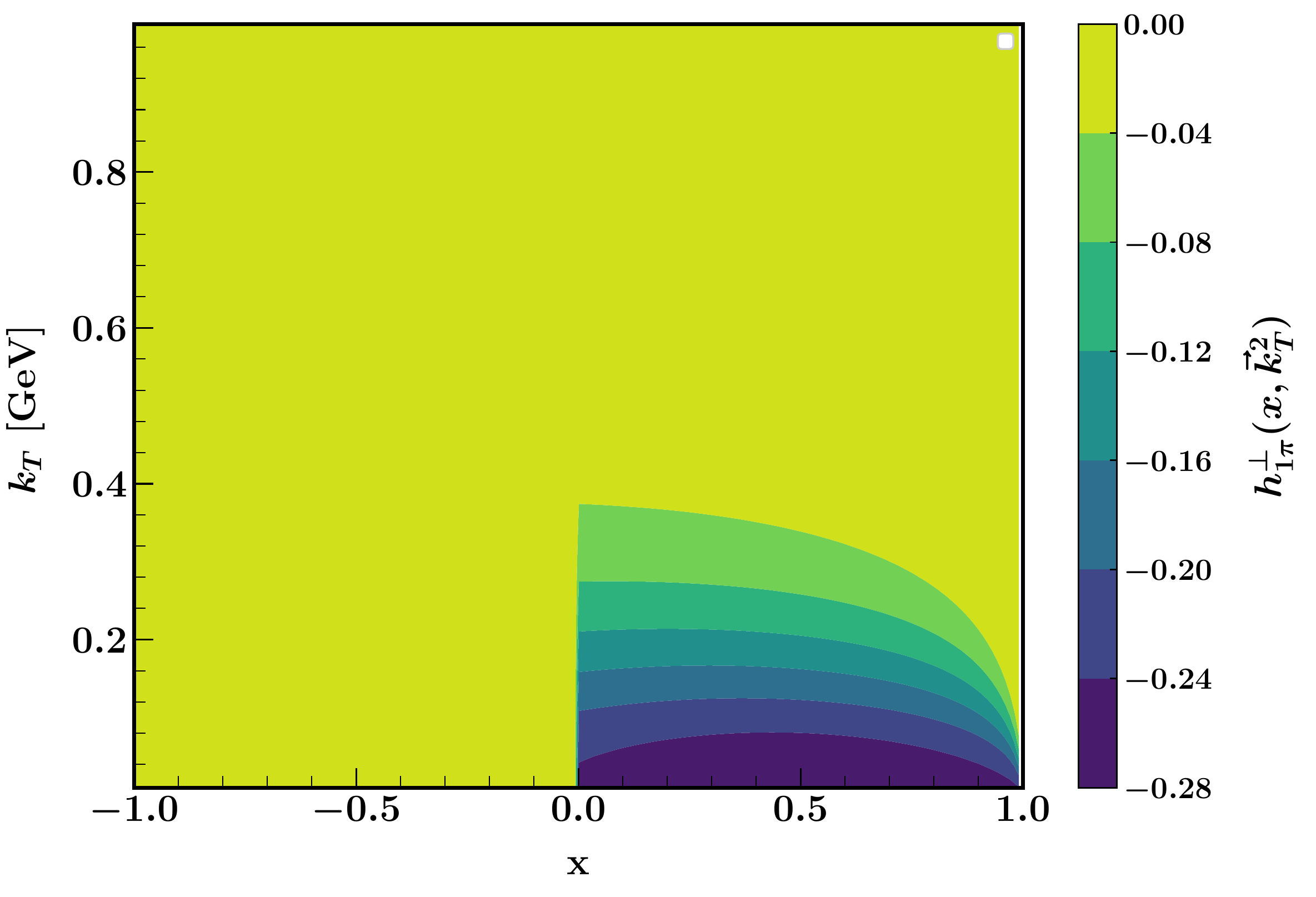}
\caption{\normalsize Quasi-TMD $h_{1\pi,Q(3)}^\perp(x,\vec k_T^2;P_z)$ as functions of x and $|\vec k_T|$ 
at different values of $P_z$ and TMD $h_{1\pi}^\perp(x,\vec k_T^2)$ as functions of x and $|\vec k_T|$.
Upper left panel: $h_{1\pi,Q(3)}^\perp(x,\vec k_T^2;P_z)$ at $P_z=1$GeV. 
Upper right panel: $h_{1\pi,Q(3)}^\perp(x,\vec k_T^2;P_z)$ at $P_z=3$GeV. 
Lower left panel: $h_{1\pi,Q(3)}^\perp(x,\vec k_T^2;P_z)$ at $P_z=5$GeV. 
Lower right panel: TMD $h_{1\pi}^\perp(x,\vec k_T^2)$ as functions of x and $|\vec k_T|$.}
\label{fig5}
\end{figure}
In the following we turn to the results of quasi-TMD $h_{1\pi,Q(3)}^\perp$ and quasi-GPD $H_{1,Q(3)}$.
In Fig.\ref{fig5}, we plot the quasi-TMD $h_{1\pi,Q(3)}^\perp(x,\vec k_T^2;P_z)$ as functions of $x$ and $k_T$ 
at different values of $P_z$ and TMD $h_{1\pi}^\perp(x,\vec k_T^2)$ as functions of $x$ and $k_T$. 
We can see from the upper left panel that the absolute value of quasi-TMD $h_{1\pi,Q(3)}^\perp(x,\vec k_T^2;P_z)$ 
becomes larger as $k_T$ close to zero. In some region with negative $x$, the corresponding $h_{1\pi}^\perp(x,\vec k_T^2)$ has nonvanishing values. 
As $x$ increases from $-0.3$ to $1$, the resulting $h_{1\pi}^\perp(x,\vec k_T^2)$ absolute value gets larger first and then becomes smaller. 
For a nonvanishing $h_{1\pi}^\perp(x,\vec k_T^2)$ result, the allowable maximum value of $k_T$ is around $0.38$. 
Comparing the three panels in Fig.\ref{fig5} depicting the quasi-TMD $h_{1\pi,Q(3)}^\perp(x,\vec k_T^2;P_z)$ with $P_z=1,3,5$ GeV, 
we can find that as the $P_z$ value increases, the minimum of $x$ inside the contours becomes larger. 
In the lower left panel with $P_z=5$ GeV, the quasi-TMD $h_{1\pi,Q(3)}^\perp(x,\vec k_T^2;P_z)$ result 
stay the same at a fixed $k_T$ value except the case $x$ close to 0 or 1.
The TMD $h_{1\pi}^\perp(x,\vec k_T^2)$ as a negative function of $x$ and $k_T$ is shown in the lower right panel of Fig.\ref{fig5}. 
The absolute value of $h_{1\pi}^\perp(x,\vec k_T^2)$ can reach 0.28 in almost all the $x$ range at a very small value of $k_T$. 

\begin{figure}[htp]
\centering
\includegraphics[height=3.5cm,width=5cm]{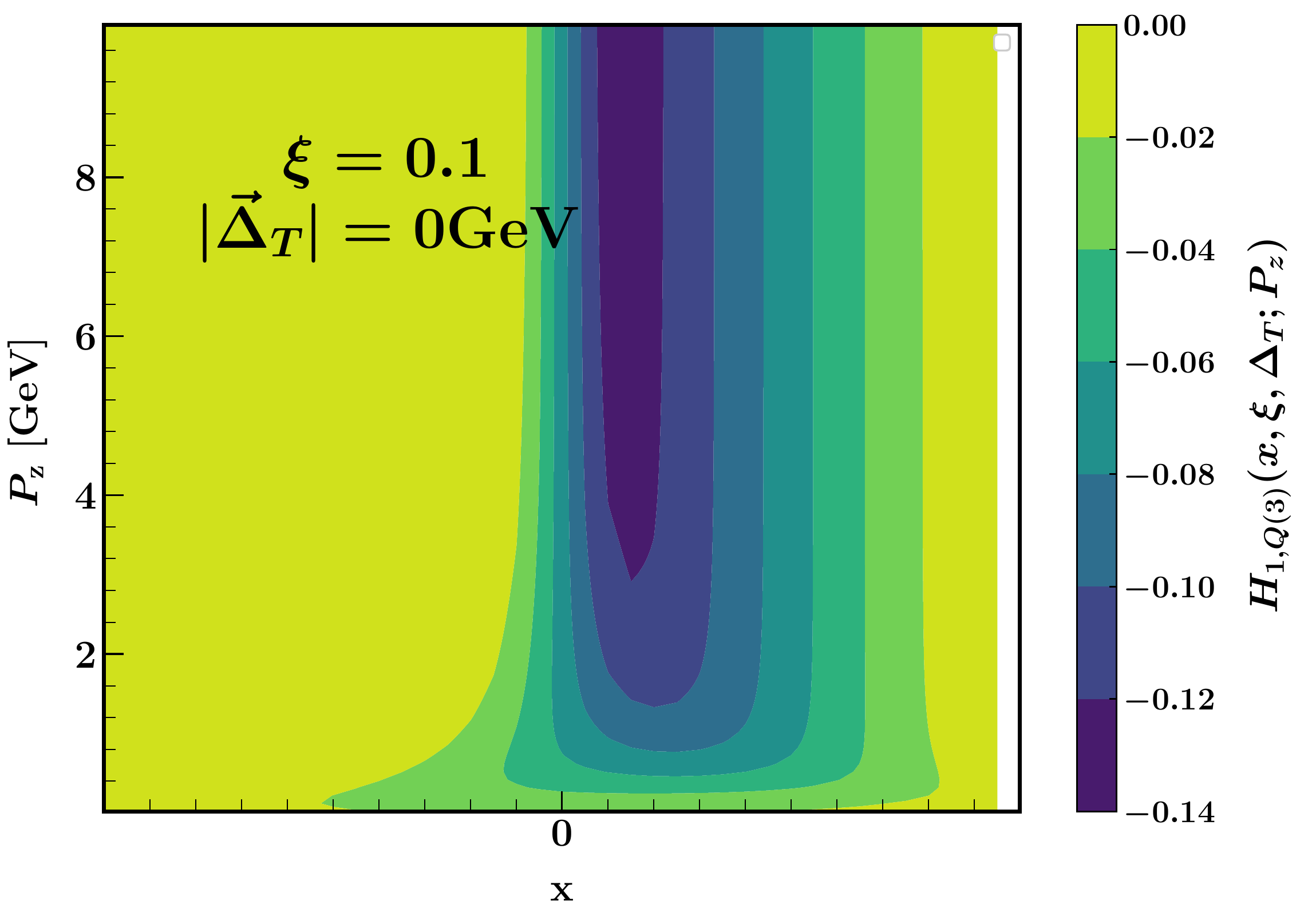}
\includegraphics[height=3.5cm,width=5cm]{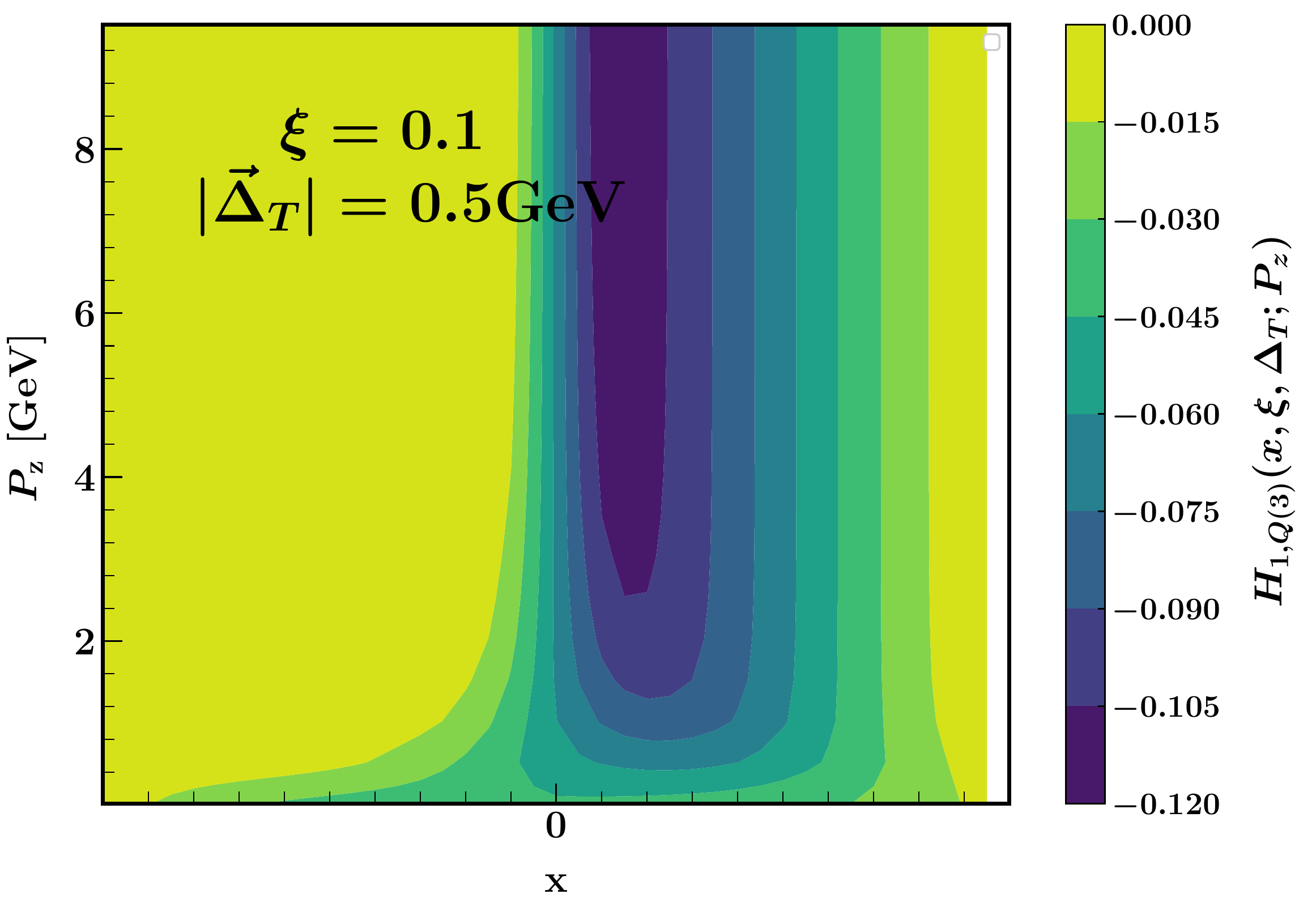}\\
\includegraphics[height=3.5cm,width=5cm]{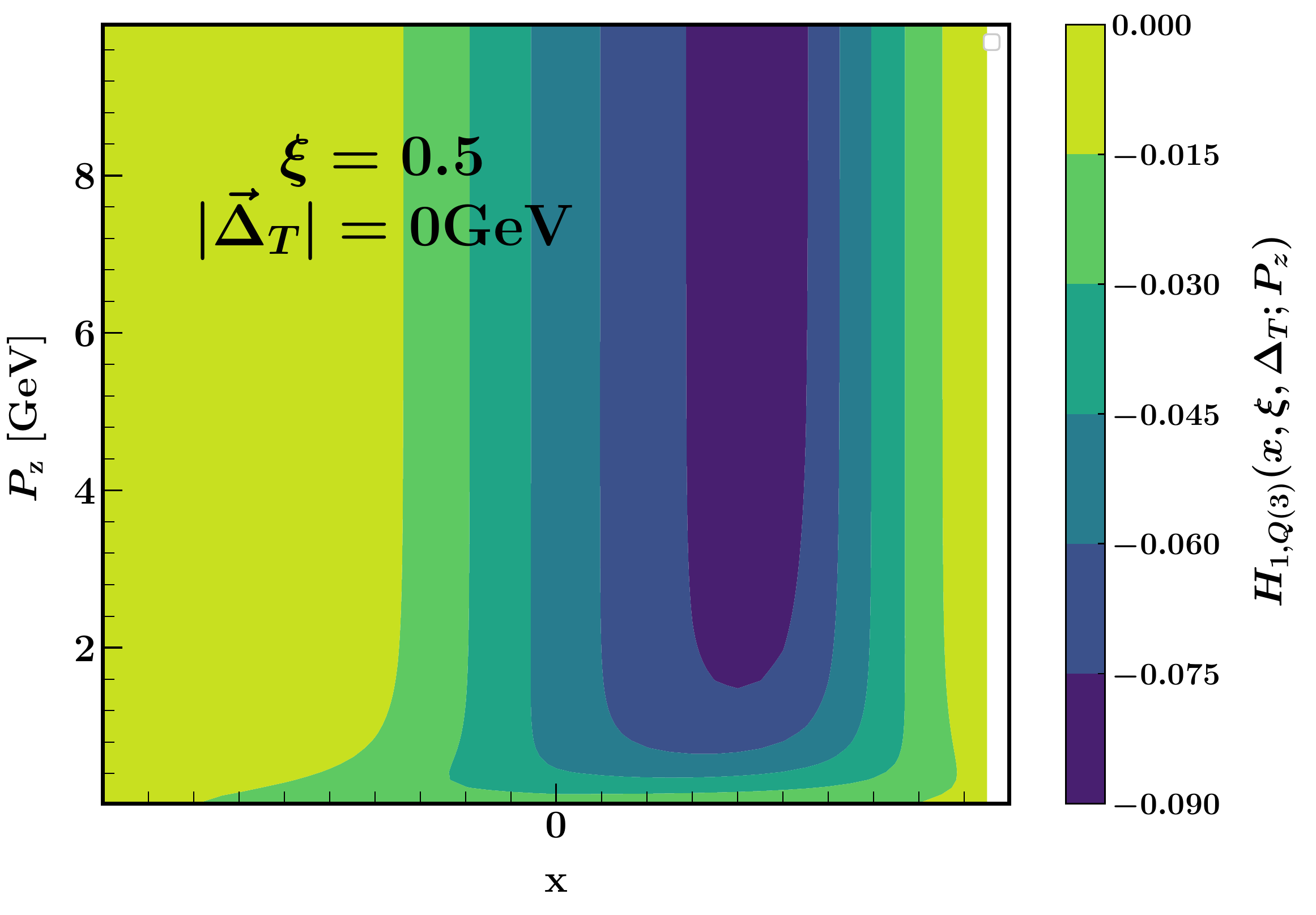}
\includegraphics[height=3.5cm,width=5cm]{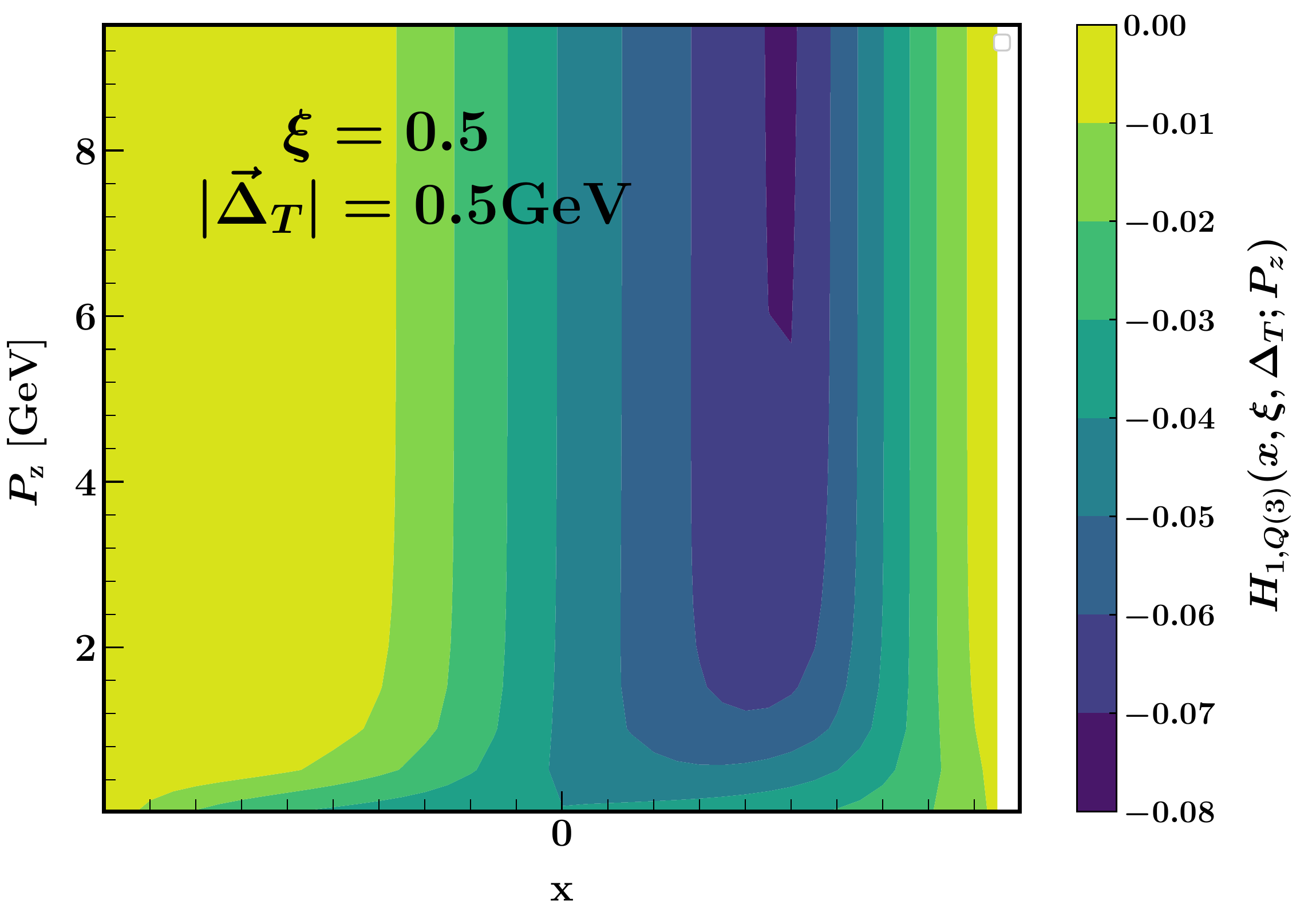}
\caption{\normalsize Quasi-GPD $H_{1,Q(3)}(x,\xi,\vec\Delta_T^2;P_z)$ as functions of $x$ and $P_z$ for different $\xi$ and $|\vec\Delta_T|$ values. 
The upper and lower panels display $H_{1,Q(3)}(x,\xi,\vec\Delta_T^2;P_z)$ at $\xi=0.1$ and $0.5$, respectively. 
The left and right panels show $H_{1,Q(3)}(x,\xi,\vec\Delta_T^2;P_z)$ at $|\vec\Delta_T|=0$ and $0.5$, respectively.}
\label{fig6}
\end{figure}
Finally, we display the quasi-GPD $H_{1,Q(3)}(x,\xi,\vec\Delta_T^2;P_z)$ as functions of $x$ and $P_z$ for different $\xi$ and $|\vec\Delta_T|$ values in Fig.\ref{fig6}. 
This quasi-GPD is negative. In the upper left panel, it is desired to mention that when $P_z > 3$ GeV, 
the quasi-GPD $H_{1,Q(3)}(x,\xi,\vec\Delta_T^2;P_z)$ result hardly depends on the value of $P_z$.
After comparing the two upper panels, we can acquire that the allowable range of $x$ for the nonvanishing quasi-GPD 
$H_{1,Q(3)}(x,\xi,\vec\Delta_T^2;P_z)$ becomes larger when $|\vec\Delta_T|$ slightly increases from zero. 
Two panels have the very similar contour shape. When $\xi=0.5$, the quasi-GPD $H_{1,Q(3)}(x,\xi,\vec\Delta_T^2;P_z)$ result 
is shown in two lower panels of Fig.\ref{fig6}. The absolute values of $H_{1,Q(3)}(x,\xi,\vec\Delta_T^2;P_z)$ in two lower panels 
are smaller than those in two upper panels. We also find that when $P_z > 3$ GeV, 
the quasi-GPD $H_{1,Q(3)}(x,\xi,\vec\Delta_T^2;P_z)$ result hardly depends on the value of $P_z$.

\section{Conclusion}
\label{V}

In this paper we have computed four T-odd GTMDs, quasi-TMD and quasi-GPD in a scalar spectator model.
We have present the results for four T-odd GTMDs. 
To get nonzero results for these functions requires considering at least one-loop corrections that include effects from the Wilson line. 
We have studied the relation of GTMDs for different values of skewness $\xi$ defined as the longitudinal momentum transferred to the proton 
and the total momentum transferred to the proton $|\vec\Delta_T|$. 
We found only in $x > \xi$ region, the T-odd GTMDs has nonvanishing value. 
Generally, the four T-odd GTMDs are negative in $x$-$k_T$ space. 
However, the $\tilde{G}_1^o(x,\xi,\vec k_T^2,\vec\Delta_T^2)$ results in certain parameter space are positive. 
Note that the T-odd GTMD $H_1^{\Delta,o}(x,\xi,\vec k_T^2,\vec\Delta_T^2)$ becomes zero when $|\vec\Delta_T|=0$. 
We have also considered the distributions of quasi-TMD and quasi-GPD. 
For the contours of quasi-TMD, we can find that as the $P_z$ value increases, the minimum of $x$ inside the contours becomes larger.
We also find that when $P_z > 3$ GeV, the quasi-GPD $H_{1,Q(3)}(x,\xi,\vec\Delta_T^2;P_z)$ result hardly depends on the value of $P_z$.

\begin{acknowledgments}
Hao Sun is supported by the National Natural Science Foundation of China (Grant No.11675033) and by the Fundamental Research Funds for the Central Universities (Grant No. DUT18LK27).
\end{acknowledgments}

\bibliography{v1}
\end{document}